# Normative atlases of neuroelectric brain activity and connectivity from a large human cohort

**Don Krieger[1], Paul Shepard[2], David O. Okonkwo[1]**

[1]Department of Neurological Surgery, University of Pittsburgh
[2]Department of Physics and Astronomy, University of Pittsburgh

Corresponding author:
e-mail: kriegerd@upmc.edu

Contributions:
Don Krieger participated in all aspects of the effort.
Paul Shepard participated in conceptualization, formal analysis, methodology, validation, writing and review.
David Okonkwo participated in conceptualization, funding acquisition, methodology, project administration, resources, supervision, validation, writing and review.

# Abstract

Magnetoencephalographic (MEG) recordings from a large normative cohort (*n* = 619) were processed to extract measures of regional neuroelectric activity and connectivity. The overall objective of the effort was to use these measures to identify normative human neuroelectric brain function. The aims were (a) to identify and measure the values and range of those neuroelectric properties which are common to the cohort, (b) to identify and measure the values and range of those neuroelectric properties which distinguish one individual from another, and (c) to identify relationships of the measures to properties of the individual, e.g. sex, biological age, and sleep symptoms. It is hoped that comparison of the resultant established norms to measures from recordings of symptomatic individuals will enable advances is our understanding of pathology.

MEG recordings during resting and task conditions were provided by The Cambridge (UK) Centre for Ageing and Neuroscience Stage 2 cohort study. Referee consensus processing was used to localize and validate neuroelectric currents, $p < 10^{-12}$ for each, $p < 10^{-4}$ for each when corrected for multiple comparisons. Comparisons of regional activity and connectivity within-subjects produced profuse reliable measures detailing differences between individuals, $p < 10^{-8}$ for each comparison, $p < 0.005$ for each when corrected for a total of $5 \times 10^5$ comparisons. Cohort-wide regional comparisons ($p < 10^{-8}$ for each) produced numerous measures which were common to the preponderance of individuals, detailing normative commonalities in brain function. Comparisons of regional gray matter (cellular) vs white matter (communication fibers) activity produced robust differences both cohort-wide and for each individual. These gray vs adjacent white matter results (1) validate that the spatial resolution of the method is better than 5 mm and (2) establish the unprecedented ability to obtain neuroelectric measures from the white matter.

The atlases derived from the results include the mean and standard deviation for each of hundreds of normative measures. These may be used to transform the same measures obtained from any individual to z-scores for statistical comparison with the norm.

# INTRODUCTION

It has been the informed expectation for a century that the keys to understanding the human brain will be found in measuring and understanding the electrical activity of neurons. Today, clinical neurophysiologists routinely measure single neurons to aide implantation of therapeutic devices deep in the brain [1]. Epileptologists use arrays of implanted "stereo EEG" electrodes and the population recordings obtained from them to diagnose and guide the treatment of intractable seizure disorders [2].

Population neuronal activity is presumed to be the basis for human behavior. Stereo EEG and comparable invasive methods produce voltage recordings with resolution of a few millimeters at best from up to a few hundred recording sites. Because the electric field interacts strongly with the conducting tissue in the brain, these measures are difficult to localize for currents at a distance from the electrodes. This problem is particularly pronounced when the recordings are made noninvasively from electrodes placed on the scalp.

Magnetoencephalography (MEG) provides an alternative noninvasive measurement approach with several advantages over scalp and even implanted EEG recordings. Electrical current flow within populations of neurons is a fundamental constituent of brain function. The magnetic fields produced by these electric currents within the brain are measured at the MEG sensor array with high fidelity. Unlike electric fields, magnetic fields do not interact with brain tissue. Therefore, in principle, the currents which are the sources of the measured magnetic field are more readily localized.

For almost all MEG studies, the neuroelectric sources of the measured magnetic fields are presumed to be due to population post-synaptic currents within the cerebral cortex [3,4]. For this reason, most source level analyses are constrained to identify neuroelectric currents in the cortex only, although there are occasional reports in which neuroelectric dipoles are localized to the white matter, e.g. [5].

In the work reported here, the referee consensus solver yields profuse measures of neuroelectric currents localized to the white matter. This can only be due to summed axial currents in bundles of nearly parallel axons. These currents and the resultant magnetic fields have been modelled [6]. The physical model has been verified and the magnetic field waveform due to a propagating action potential (AP) has been measured from the medial giant axon of the crayfish [7].

Production of a detectable magnetic field requires coincidence of many APs within a small volume of white matter, i.e. passage of a volley. Since the individual APs are of short duration, 1-4 msec, the duration of the volley must be near that to avoid overwhelming destructive interference between the magnetic fields produced by the individual APs.

Measurement of activity with duration of 5-10 msec requires sampling with band pass at 200 Hz or more. The great majority of published findings in MEG are confined to bands below 25 Hz with 1-80 Hz being the most typical pass band [8]. In many studies, the pass band is cut off 5-10 Hz below the line frequency, i.e. 5-10 Hz below 60 Hz in the United States. Hence in the majority of published MEG work, detection of neuroelectric currents within the white matter is de novo excluded by confining the measurements (1) to low frequencies and (2) to the cerebral cortex.

With the pass band open to 250 Hz and the neuroelectric current localization unrestricted, substantive difficulties remain. The measurements at the MEG sensor array are due to an unknown and almost certainly large number of neuroelectric currents. If the location of a single contributor to the MEG measurements can be identified, i.e. the location of a single neuroelectric current, then the corresponding current can be accurately computed using the Biot-Savart law [9], regardless of the number and amplitude of other currents. Hence the fundamental problem of extracting source space information from MEG recordings is deconvolution, i.e. separation of the signal from each current from that of all the others sufficiently to accurately identify the location of each current, one at a time.

In restricted special cases, the data may be manipulated so that the activity of only one or a few currents is a significant contributor to the MEG. Most commonly this is accomplished (1) by averaging multiple data segments, each synchronized to an event, e.g. a stimulus, and (2) by restricting the analysis to MEG sensors in which the signals are expected to be largest for the expected current(s). In these cases, Equivalent Current Dipole (ECD) source localization may be used. The signals in the selected sensors are fit to the forward solution for one or a few point dipole sources. This nonlinear optimization problem is typically solved using a damped gradient descent algorithm which works well typically for one source, e.g. [10,11], but may be effective for two to four sources, e.g. [12-14], so long as there is little interference in the measurements from simultaneously active currents.

For the general multiple source problem, many investigators have opted to use methods with which thousands of point sources are estimated in a single operation, e.g. MNE [15], LORETA [16]. These approaches produce estimates of activity from fixed current locations with sufficient density to ensure that no presumed true current is more than a few mm from one of them. These methods produce thousands of parameters from hundreds of data points; the problem is "poorly posed" and the solution is grossly over-fit. Because of this, localization accuracy is compromised and the ability to resolve currents which are near each other is poor. In addition, these methods are vulnerable to interference which is typically minimized using up front averaging with consequent loss of information.

ECD, MNE, LORETA, and comparable methods all fit a single model to the observed magnetic field measures. The referee consensus method avoids the fundamental problem that the number of contributors to the MEG is unknown and likely large. Instead of fitting a global model to the MEG, it uses a gradient search to identify a single current dipole at a time. It is like ECD in that it uses a search but it is different in that it fits a model to the dipole time series within a narrowly defined subspace of the MEG measures. That signal subspace is defined by a novel and tight spatial filter, i.e. a projection operator, which constrains the measures to those which closely fit the correct amplitude ratios for the signal due to a current at the search location, i.e. the forward solution.

In this way, the referee consensus method is like other methods which explicitly use spatial filters, e.g. [17-21]. Such methods avoid the generally unsolvable problems of accurately accounting for all of the information contained in the MEG recordings without knowing the number of contributors and without having sufficient measures to solve the equations. For the cited examples, all of which are beam formers, a set of filters is generated, each of which is optimized to yield source space measures from a constrained volume of the brain while minimizing interference from currents in "nulled" brain volumes.

In summary, there are four properties of the referee consensus method which distinguish it. (1) It does not use a global solution like ECD, MNE, LORETA, and others in which the MEG recordings are fit to a single source space model. Hence it avoids both the over-fitting problem and the problem that the number of contributors to the MEG recordings is unknown. (2) The method uses spatial filters which are optimized to measure differentials in the close vicinity of one test location at a time. These filters are paradoxical in that their transfer functions include zeroes either at or very near the test location.

(3) For each test location 1080 filter sets are used to compute 1080 separate "referee opinions" regards the presence of a current at the location. The signal/noise enhancement provided by using a family of filters to make each of these decisions is estimated at 50. This is the key property which enables extraction of useful information from the raw recordings without averaging. (4) Each of the 1080 filter sets produces an estimate of the waveform for the current at the test location. The signal/noise for each waveform estimate is about 5. The signal/noise enhancement obtained by combining the 1080 waveform estimates is estimated at 50. The appendices detail (I) the method and (II) its advantages in extracting and localizing neuroelectric measures. The Methods section presents the dataset and then the sequence of processing steps, primary, secondary, and tertiary.

It is the secondary processing which is central to understanding the results, particularly the definitions and normalizations for regional neuroelectric activity and regional network connectivity. It is the normalizations described in equations (1) and (3) which are the bases for the differential measures implemented in the tertiary processing steps. The subsections in the Results section mirror the subsections in the Methods section.

The paper closes with conclusions and suggested steps going forward. This last is particularly apropos given the path-finding nature of the work in this report. There are numerous alternative approaches which are yet to be explored with the profuse and reliable measures yielded by the referee consensus solver. Directions for access to the full set of those measures for the CamCAN lifespan cohort may be may be found at http://stash.osgconnect.net/+krieger/. This transformed data set is approximately three TBytes. Example files from this data set are contained in the supplementary file, S1_Tables.zip.

# Materials and Methods

Magnetoencephalographic (MEG) recordings were processed from each member of a normative cohort, n = 619, ages 18-87 [22,23]. The raw data from each subject was initially transformed to a collection of probabilistically validated neuroelectric currents, mean = 701,020 per minute per subject. Each current is 80 msec in duration and is localized in time and space with millisecond (msec) and better than 5 millimeter (mm) resolution. This primary processing step yielded profuse validated high resolution neurophysiological measures from within the brain of each subject.

These currents were then counted to produce normalized measures of activity and network connectivity for each of 158 standard regions of interest (ROIs). This data reduction step produced tonic regional measures. Each regional measure is a count of all the neuroelectric currents localized within the region over the nine-minute recording. The statistical power when comparing counts to test for regional differences is high because the count for each ROI is high.

Tests using the $\chi^2$ statistic are used throughout the report to provide confidence that comparisons do or do not show differences. In most cases, the counts are large, providing considerable statistical power. For all comparisons, the threshold p-values which are used to decide if results are significant or not is stringent. The large number of significant findings and the magnitude of the statistics recommend both (a) the validity of the results and (b) the potential for benefit from analyses which utilize the high-resolution temporal and spatial information in the neuroelectric current measures.

# CamCAN dataset

The Cambridge Centre for Ageing and Neuroscience (CamCAN) Stage 2 cohort study is a large cross-sectional adult lifespan study (ages 18-87) of the neural underpinnings of successful cognitive ageing [22,23]. The work reported here utilized the majority subset (*n*=619) of the cohort for whom high resolution (1 mm) anatomic T1-weighted MR imaging and MEG recordings were available.

MR imaging was obtained on all subjects at a single site using a 3T Siemens TIM Trio scanner with 32-channel head coil. The T1-weighted imaging was obtained using the MPRAGE sequence. The field of view for these scans was 256 x 240 x 192 at 1 mm resolution.

MEG recordings were collected at a single site using a 306-channel VectorView MEG system (Elekta Neuromag, Helsinki). The data were sampled at 1 KHz with anti-aliasing low-pass filter at 330 Hz and high-pass filter at 0.03 Hz. Continuous head position measures were enabled throughout the recordings. All recordings were obtained with the subject sitting up.

560 seconds were recorded continuously with eyes closed resting [22]. 560 seconds were recorded continuously during performance of a sensorimotor task. Subjects detected visual and auditory stimuli and responded to detection of each with a button press with the right index finger. The stimuli were two circular checkerboards presented simultaneously to the left and right of a central fixation cross, 34 msec duration, and a binaural tone of 300 msec duration. The tone was at 300, 600, or 1200 Hz in equal numbers with the order randomized. 121 trials were presented with simultaneous visual and auditory stimulation. Eight trials were randomly intermixed in which one stimulus was presented at a time, four visual and four auditory. This was done to discourage dependence on one stimulus modality only. The average inter-trial interval was approximately 4.3 seconds

120 seconds were recorded continuously during passive attendance to the same stimuli as those presented in the sensorimotor task [22]. Here the visual and auditory stimuli were presented singly rather than simultaneously and no response was required. The average inter-trial interval was 1 per second.

Anxiety and depression scores were obtained from the Hospital Anxiety and Depression Scale (HADS) [23,24]. Of the 619 subjects included in the study, supra threshold scores were obtained for 41 subjects for anxiety and 13 for depression. The elevated number for anxiety scores perhaps reflects testing induced anxiety. The Addenbooke Cognitive Examination Revised (ACE-R) [25] total score was used as a measure of cognitive function and the Pittsburgh Sleep Quality Index (PSQI) [26] was used as a measure of sleep quality.

The data processing pipeline is detailed below and is schematized in Table 1.

| | | | | | | |
|---|---|---|---|---|---|---|
| MEG Recordings | 1° ➔➔➔ | Neuroelectric currents: 80 msec duration ≈10,000 per second $p < 10^{-12}$ for each | 2° ➔➔➔ | Normalized regional activity and connectivity density | 3° ➔➔➔ | Individual and cohort-wide properties $p < 10^{-8}$ for each |
| | Referee Consensus Solver | | Regional segmentation and counting | | Comparison of regional density differences | |

Table 1. Processing pipeline summary. Primary (1°), secondary (2°), and tertiary (3°) processing are detailed in the sections below.

# Primary Processing

## MRI processing, MEG filtering, and continuous head positioning

The high resolution T1-weighted MRI scan was processed with Freesurfer, version 5.3 [27,28]. Freesurfer is a segmentation package which automatically and reliably identifies brain regions. The 3-dimensional coordinates of the extent of the brain volume and 232 standardized regions of interest (ROI's) were identified. Figure 1 shows the expected descending relationship between brain volume and age.

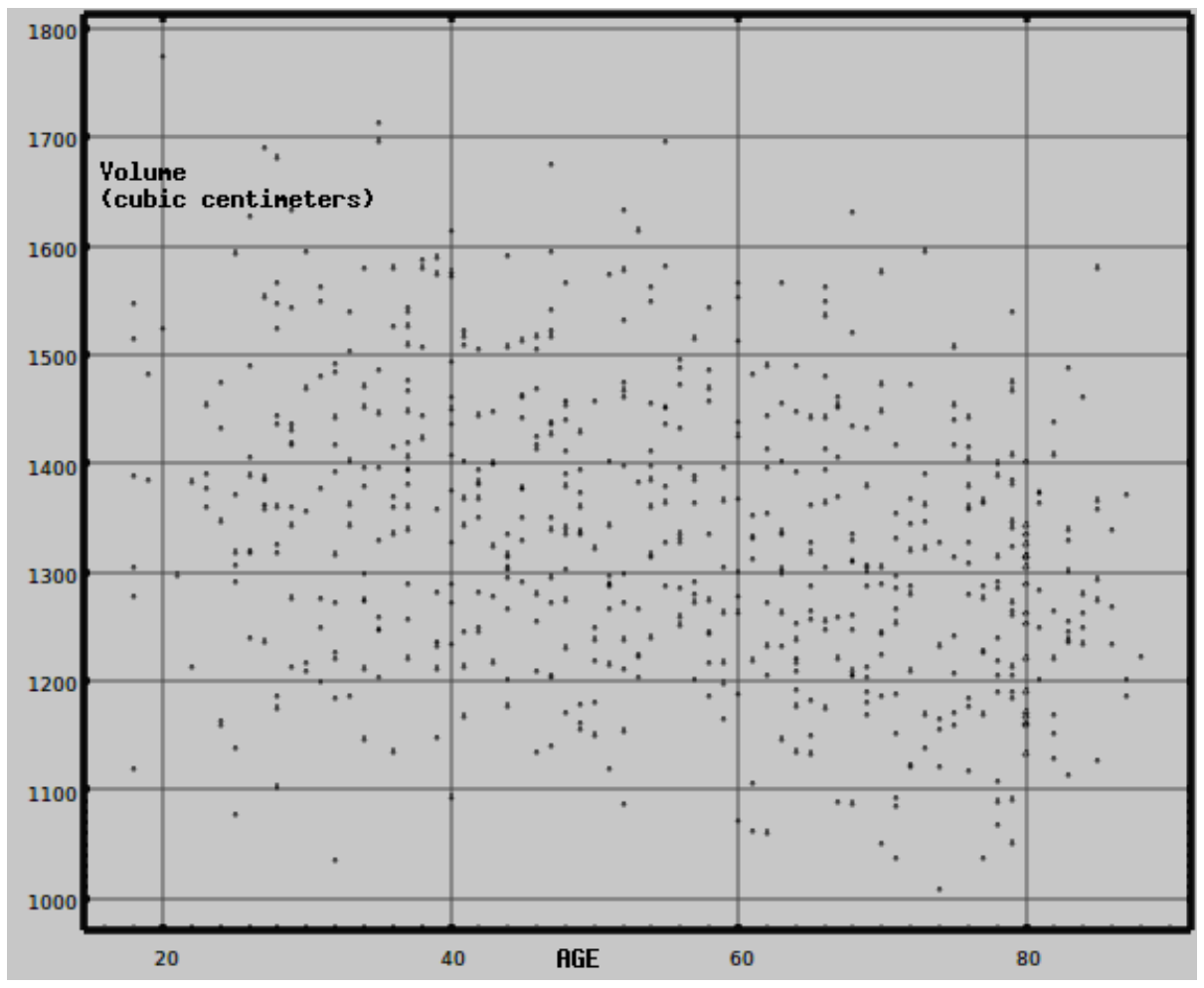

Figure 1. The brain volume for each subject was obtained using Freesurfer (see text). Brain volume for this normative cohort shows the expected decrease with age [29] beginning at about age 60. Over the full age range, the Pearson's correlation between brain volume and age is -0.295, $df = 617$, $p < 10^{-18}$

The subject's head position within the MEG scanner was manually coregistered to the TI-weighted MRI scan using Elekta's Mrilab visualization tool. The coordinates of the center point of a sphere most nearly approximating the brain was identified. Using the brain segmentation provided by Freesurfer, a set of non-overlapping 8x8x8 mm voxels was identified which cover the brain volume excluding the sphere at the center of the volume with 30 mm radius.

Continuous head positioning measures were extracted using Elekta's MaxFilter tool [30]. The MEG channels were each filtered with high and low pass at 10 and 250 Hz, 5 Hz roll-off, using MNE tools [31]. Note that the 10 Hz high pass filtering effectively demeans each channel. For each 1.24 second data segment, mains noise was removed at 50, 100, 150, 200, and 250 Hz using polynomial synchronous noise removal [32]. No other preprocessing was applied and no data segments were excluded by manual artifact identification. Instead, artifact rejection relied upon the solver's inherent failure to identify validated neuroelectric currents when presented with excessively noisy data.

## MEG processing: Identification of validated neuroelectric currents

The coregistration of the MEG sensor array with the location of the subject's head and brain was corrected once per second using the continuous head positioning information. This correction was applied to the forward solution used by the solver.

The forward solution is the mathematical relationship between a putative electric current within the brain and the resultant magnetic field measurements at the sensor array. It models the brain as a uniformly conducting sphere [9]. Currents within 30 mm of the center of the sphere are nearly undetectable and the mathematical formulation for the forward solution is poorly behaved for this volume; hence it was excluded from the search. The intersection of this region with an MRI slice is shown in Figure 2. Note that the excluded volume typically includes the posterior thalamus, the posterior commissure, and much of the midbrain (not shown in the figure).

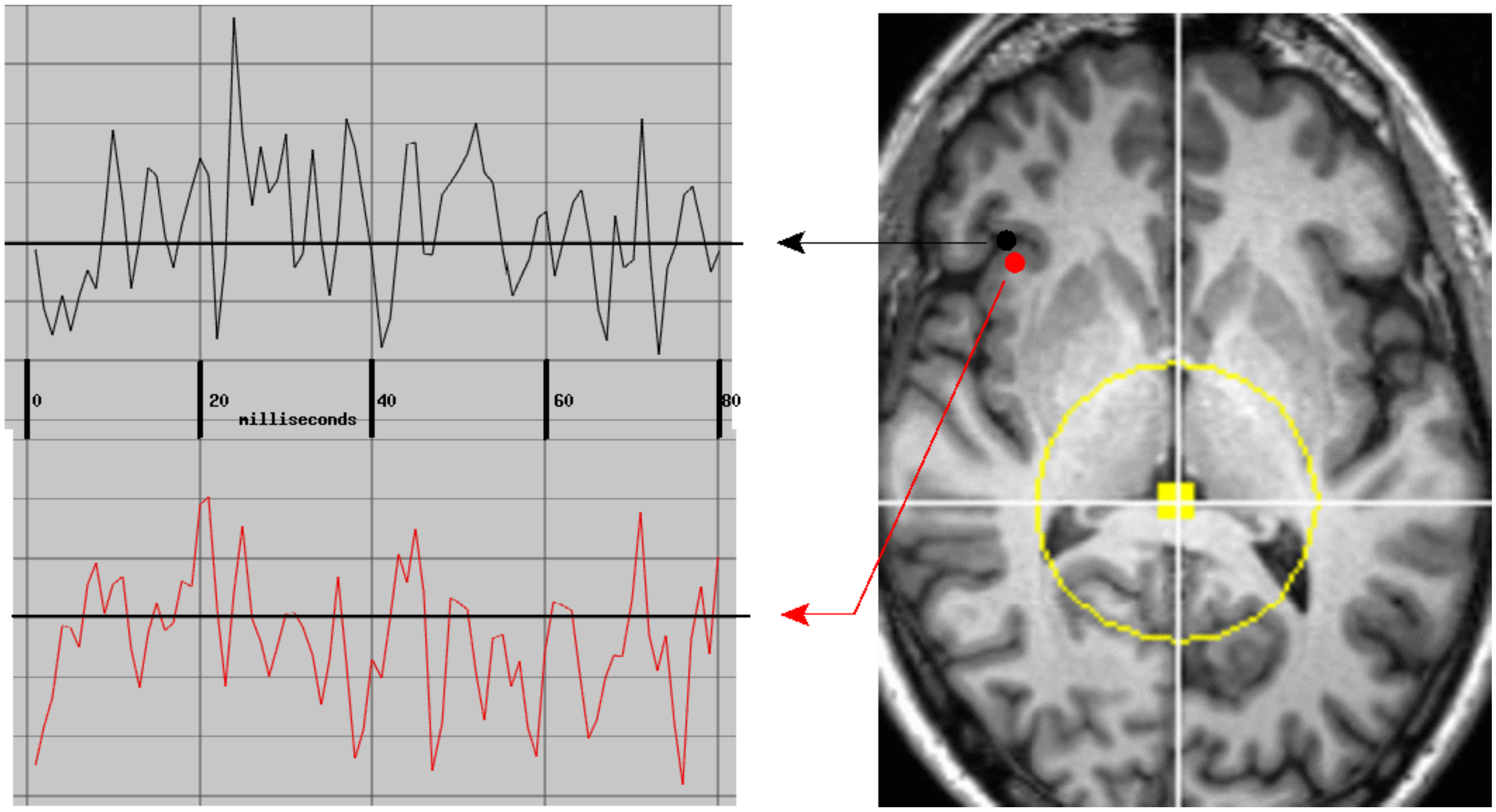

Figure 2. Two typical simultaneously active neuroelectric currents were identified and validated by the referee consensus solver, $p < 10^{-12}$ for each, i.e. $p < 10^{-4}$ for each when corrected for multiple comparisons. Each waveform has duration of 80 msec sampled at 1000 Hz. The bandpass is 10 – 250 Hz. The currents are 5 mm apart with zero-lag cross-correlation of 0.157, $df = 80$, $p = 0.16$. The yellow dot and circle delineate the region near the center of the head which is excluded from the search for neuroelectric currents. See text for details.

The referee consensus solver [33,34] was applied to the continuous recordings to identify and validate neuroelectric currents throughout the brain. The search of the brain volume was conducted for one 80-msec data segment at a time. The search progressed through the 560 second data stream in 40 msec steps, i.e. 25 steps/second. The number of searches to cover the volume of the brain for an entire recording session was less than $10^8$. The choice of $10^{-12}$ as the threshold for acceptance yields a threshold corrected for multiple comparisons of $p < 10^{-4}$ for each validated neuroelectric current. This provides confidence that almost all identified and validated currents are real.

## Secondary processing

### Regional counts

Rather than pursuing analyses which capitalize on the high time resolution of the current waveforms, it was the stability of the measures over time which was emphasized in this work. Counting for several minutes of data was used to obtain measures of tonic activity, i.e. the number of identified currents per unit volume. Counts were obtained for standardized regions of interest to enable volumetric comparisons between individuals. This also enables comparisons of MEG-derived results with results derived from other functional imaging modalities, e.g. fMRI.

As described above, freesurfer [27] was used to identify 232 standard brain regions of interest (ROIs). Both regional neuroelectric activity and connectivity were normalized to enable comparisons between ROIs and between activity and connectivity. The ROIs vary widely in volume. This approach therefore surveys regional measures with a wide range of statistical power. That information will be useful in refining the approach.

For each cortical ROI, freesurfer identifies the adjacent white matter ROI with rim up to 5 mm thick. It also identifies a border of tissue 1 mm thick dividing the cortex and white matter. There are 68 standard ROI triples. Because their volumes are so small, the 68 border ROIs were excluded from most of the analyses. The optic chiasm and five subdivisions of the corpus callosum ROIs were also excluded due to their small volumes. This left a total of 158 ROIs for which regional counts were analyzed, 68 cortical, 68 adjacent white matter, and 22 other ROIs including unsegmented white matter, cerebellar ROIs, sub-cortical ROIs, the corpus callosum, and the brain stem.

## Regional neuroelectric activity

The validated neuroelectric currents identified within each ROI were counted and normalized to a current density, $\rho_{roi}$, as shown in equation (1). The purpose of this normalization was to enable comparisons of an ROI with or between individuals, during different states, at different times, or comparisons of one ROI with another. The normalization was defined so that $\rho_{roi} = 1.0$ for all ROIs if the neuroelectric currents were uniformly distributed throughout the brain. In that isotropic case, the regional count fraction was always equal to the regional volume fraction and the null hypothesis was true for all comparisons.

$$\rho_{roi} = (count_{roi}/count_{total}) \div (vol_{roi}/vol_{total}) \quad (1)$$

$count_{roi}$ is the count of the currents found inside the ROI, $count_{total}$ is the count of all the currents found inside the brain, $vol_{roi}$ is the volume of the ROI, and $vol_{total}$ is the total volume of the brain.

Thus $\rho_{roi}$ is a dimensionless "density." It is the ratio of the ROI's fractional count and fractional volume. Note that dividing the counts for an ROI by the total count normalizes the "density" for variations due to both data quality and record length. The normalization for data quality is particularly important since poor data quality markedly reduces the yield of the solver, e.g. Figure 5.

The count of validated neuroelectric currents is high because the method is effective when applied to the raw data stream. The false identification rate is low because the validation threshold is stringent, $p < 10^{-12}$, i.e. $p < 10^{-4}$ per individual identified current when corrected for multiple comparisons. These characteristics produce ample statistical power to identify differences between regional density measures, e.g. for a cortical ROI vs the adjacent white matter ROI, or for an ROI during rest vs the same ROI during task. The activity densities are reduced in deep structures and in the cerebellum. This is likely due to reduced sensitivity at the MEG sensor array as the distance to the magnetic field source increases. However, the numbers are ample even in the brainstem to see significant differential activity, e.g. when comparing rest and task.

## Regional network connectivity

A normalized connectivity measure was computed for each ROI, $q_{roi}$ as shown in equation (3) below. As was described above for regional neuroelectric activity, the purpose of normalization is to enable comparisons of an ROI with or between individuals, during different states, at different times, or comparisons of one ROI with another. The basis for the connectivity measure is a count of pairs of simultaneously occurring neuroelectric currents. The normalization is defined so that $q_{roi} = 1.0$ for all ROIs if both legs of all pairs are uniformly distributed throughout the brain. In that isotropic case, the null hypothesis is true for all comparisons and, as shown in equation (2b), the ROIs' count fractions are always equal to a product of two volume fractions: (1) the volume fraction of the ROI within which one leg of the pair is constrained multiplied by (2) the volume fraction of the region within which the other leg is constrained.

The equivalence between count fraction and a product of volume fractions is clarified by an example. Consider a large number of simultaneously active pairs with both legs distributed uniformly throughout the brain, $count_{total}$. For each pair, we identify one leg as #1 and the other as #2. Suppose we have an ROI whose volume is 0.1 of the brain and we count the pairs with one leg inside the ROI and the other leg outside, i.e. in the remaining 0.9 of the brain. Since the distribution of both legs of the pair is uniform, 0.1 of both legs are inside the ROI and 0.9 of both are outside. For the 0.1 with leg #1 inside the ROI, 0.9 of those pairs have leg #2 outside and are included in $count_{roi}$. In addition for the 0.9 with leg #1 outside, 0.1 of those pairs have leg #2 inside and are also included in $count_{roi}$. Hence $count_{roi} = count_{total} \times [0.1 \times 0.9] \times 2$. In general then,

$$count_{roi} = count_{total} \times [(vol_{roi}/vol_{total}) \times ((vol_{total} - vol_{roi})/vol_{total})] \times 2 \qquad (2a)$$

We divide both sides of equation (2a) by $count_{total}$ to get the expression for the ROI count fraction,

$$count_{roi}/\ count_{total} = [(vol_{roi}/vol_{total}) \times ((vol_{total} - vol_{roi})/vol_{total})] \times 2 \qquad (2b)$$

The equivalence between ROI count fraction, left side of equation 2b, and the product of volume fractions on the right side of the equation enable defining $q_{roi}$ to be the analogue of $\rho_{roi}$ for pairs as shown in equation (3).

$$q_{roi} = (count_{roi}/count_{total}) \div [(vol_{roi}/vol_{total}) \times ((vol_{total} - vol_{roi})/vol_{total})] \times 2, \text{ where} \qquad (3)$$

$count_{roi}$ is the number of simultaneously occurring pairs for which one member of the pair is inside the ROI and the other member is outside.

For a cortical or adjacent white matter ROI, the outside member of the pair is constrained to fall outside both the cortical ROI and the adjacent white matter. This also applies to pair counts for either the cerebellar gray or white matter ROI's. This constraint is included to exclude counting pairs which span only local connections. $count_{total}$ is the count of all such pairs found for all ROIs, $vol_{roi}$ is the volume of the ROI, and $vol_{total}$ is the total volume of the brain.

## Coincidence of neuroelectric current pairs

The average number of simultaneously active (coincident) currents validated at $10^{-12}$ over the entire cohort is approximately 440. For an interval with 440 coincident currents, the total number of possible pairs is equal to the combinations of 440 currents taken 2 at a time, C(440,2) = (440 x 439)/2. The subset of those pairs which are likely functionally coupled was selected by identifying the pairs which occurred much more often than by chance, $p < 10^{-8}$ for each.

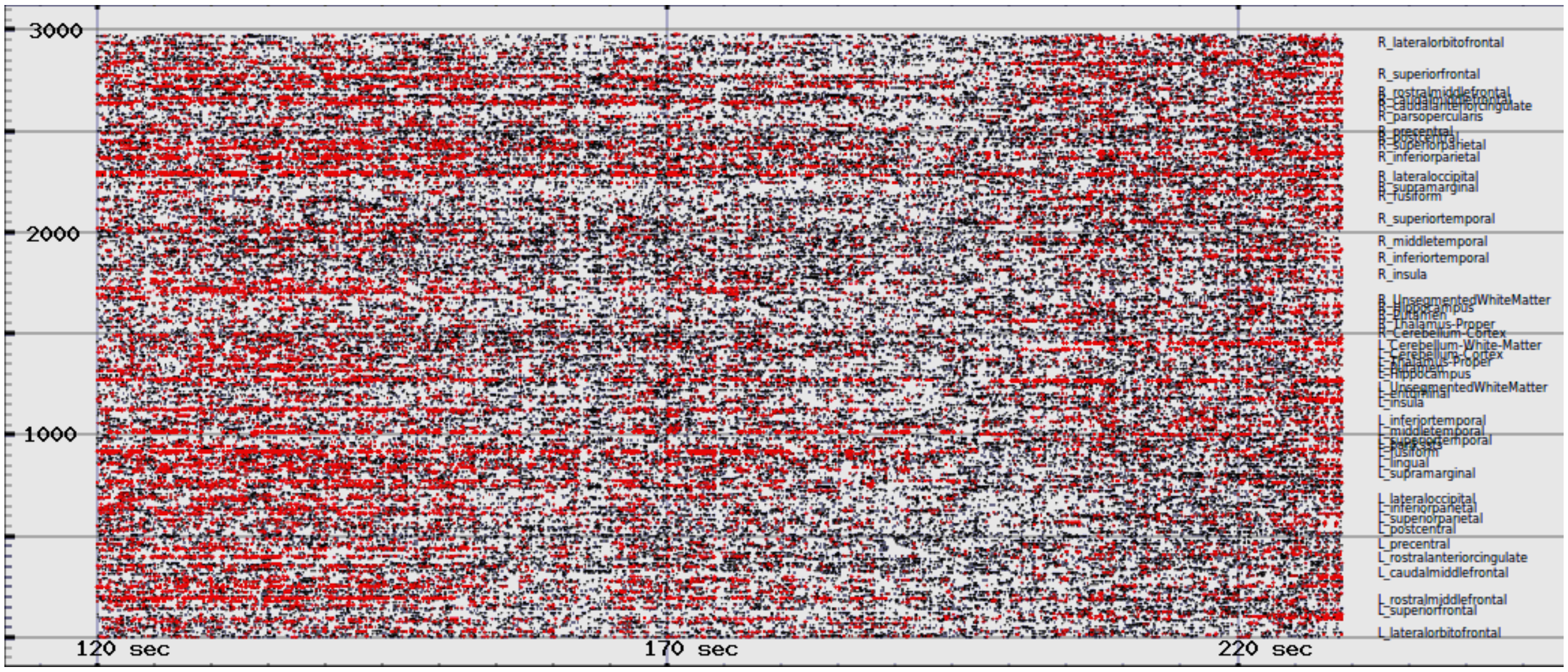

Figure 3. Two minutes from the resting recordings of a single subject are shown. The 3000 brain locations at which the largest number of validated currents occurred were identified. The names of most of the ROIs into which these fall are listed at the right. The time at which each current occurred is plotted as a black dot. The times at which a simultaneously active pair of validated currents occurred are plotted as red dots.

Computing this probability is conceptualized and implemented as follows. Consider a recording with 400 second duration. The solver proceeds through the recording in 40 msec steps, i.e. 25 per second. Hence there are 400x25=10,000 intervals. Suppose that a current was identified at location #1 within the brain m times and at location #2 n times. What is the probability that the currents found at both locations will coincide at least $\ell$ times?

We reformulate this question as one of sampling with replacement. We place 10,000 balls which represent the intervals in a bowl of which m are red and the rest are white. Each of the m red balls represents an interval during which a current was identified at location #1. We now draw a ball from the bowl n times and each time we put it back when we are done. Each of the n draws represents an interval during which a current was identified at location #2. We return the ball to the bowl each time because each draw represents a different interval. What is the chance that we will draw at least $\ell$ red balls from the bowl? On any draw, the chance that we will draw a red ball is m/10000. The chance that we will draw <u>exactly</u> $\ell$ red balls is $[m/10000]^{\ell}$ x $C(n,\ell)$. Finally, the chance that we will draw <u>at least</u> $\ell$ red balls is $\sum_{\mathbf{k}=\ell,\mathbf{n}}\{[\mathbf{m}/\mathbf{10000}]^{\mathbf{k}}\ \mathbf{x}\ \mathbf{C}(\mathbf{n},\mathbf{k})\}$

CDFBIN [35] was used for the calculation. The threshold probability for accepting a pair for inclusion was $10^{-8}$. This value was chosen as follows. The number of neuroelectric currents found within each 1 $mm^3$ brain voxel was counted and sorted. The 3000 locations with the highest counts were selected as the individual locations which would be included in simultaneously occurring pairs as shown in figure 3. This limited the counts to (3000x2999)/2 = 4,498,500 possible location pairs, a manageable number both computationally and probabilistically. With approximately 5 million pairs to be tested, the chance that even a single pair would produce a probability less than $10^{-8}$ by chance is 0.05 .

# Tertiary processing - comparisons

For each ROI the secondary processing provides both a count and the corresponding normalized density measure. When comparing activity or connectivity for one ROI for a single subject, it is the counts which are compared using the $\chi^2$ statistic, e.g. equation (3).

$$\chi^2 = \sum_{i=1,2}\left( {(\text{observed}_i - \text{expected}_i)^2}\Big/{\text{expected}_i} \right) \qquad (3)$$

The calculations used to normalize the count to a density measure as formulated in equations (1) and (2) are also used to compute the expected counts. To control for false positive findings due to multiple comparisons, $p < 10^{-8}$ was used as the threshold for significance. This is conservatively corrected to $p < 10^{-2}$ for each test since there are less than $10^6$ comparisons (619 subjects x 158 ROIs/subject) for each of seven within-individuals comparisons. It is noteworthy that because the counts are so high for each subject's ROIs, there is considerable statistical power for identifying differential regional activity and connectivity.

When comparing activity or connectivity of each ROI for the entire cohort, the differences between the *means* of the density measures may be compared using Welch's t-statistic. A more conservative approach was used. The comparison for each ROI for each individual provided a measure of confidence for a difference, e.g. between the regional activity during rest vs task. The $\chi^2$ statistic was used to determine if one side of the comparison was greater than the other in the preponderance of subjects. For this test, the contribution to the $\chi^2$ statistic for each ROI is the same for each subject. Hence the test is not vulnerable to the effects of outliers. For a group comparison, there is one test per ROI, i.e. 158 tests. For these, $p < 10^{-8}$ was used as the threshold for significance. Hence the chance that even a single test would appear significant was $p < 10^{-5}$.

## Rest vs task activity

For each subject, the count for each ROI during rest was compared with the count during the task. As stated above, $p < 10^{-8}$ was the threshold for acceptance of a significant difference, i.e. $p < 10^{-2}$ for each test corrected for multiple comparisons.

For each ROI for the entire cohort, the number of individuals who demonstrated a significant difference with rest > task was compared with those with rest < task. $p < 10^{-8}$ was used as the threshold for a significant preponderance, i.e. $p < 10^{-5}$ for each test corrected for multiple comparisons.

## Cortex vs adjacent white matter activity

Each cortical ROI was paired with the adjacent white matter ROI. The white matter ROI extends into the cortical gyri and includes a white matter rim whose thickness is limited to 5 mm. The comparison for each ROI pair between cortex and white matter was computed separately for each individual. For each comparison, the counts for the two ROIs were compared using $\chi^2$ with $p < 10^{-8}$ used as the threshold for significance for each test, i.e. $p < 10^{-2}$ corrected for multiple comparisons.

For each ROI, the following comparison was carried out for the cohort as a whole. The number of subjects who demonstrated a significant difference with cortex > adjacent white matter was compared with the number with cortex < adjacent white matter. $p < 10^{-8}$ was used as the threshold for a significant preponderance, i.e. $p < 10^{-5}$ for each test corrected for multiple comparisons.

## Activity vs network connectivity

The activity and connectivity density measures were computed with the same normalization and scaling. Hence they may be directly compared. The comparison for each ROI between activity and network connectivity was computed separately for each individual. For each comparison, the counts for the two measures were compared using the $\chi^2$ statistic. As stated above, $p < 10^{-8}$ was the threshold for acceptance of a significant difference, i.e. $p < 10^{-2}$ for each test corrected for multiple comparisons.

For each ROI, the following comparison was carried out for the cohort as a whole. The number of individuals who demonstrated a significant difference with connectivity > activity was compared with the number with connectivity < activity. $p < 10^{-8}$ was used as the threshold for a significant preponderance, i.e. $p < 10^{-5}$ for each test corrected for multiple comparisons.

### Global activity patterns

The secondary processing described above provided a set of 158 activity density measures for each subject. Hence each subject may be conceived as a point in a 158-dimensional space. Principle components (PC) analysis [36,37] was then applied to reduce this number, 158, to 20 linear combinations of these measures which are orthogonal to each other, i.e. uncorrelated.

The subset of 20 PCs which capture the maximum variance in the original data were selected. This reduction in the number of measures was necessary to enable practical use of all subsets regression [38] as described below. It also favors inclusion of information from each individual which is representative of the cohort.

The "first" 20 PCs accounted for 38.6% of the total variance. The 1st PC accounted for 12.8 times its fair share of the variance; the 20th accounted for 1.8 times its fair share.

All subsets linear regression was applied to identify linear combinations of these 20 PCs which were optimally correlated with either (a) subject age (b) subject sex, (c) anxiety, (d) depression. Anxiety and depression symptom scores were used from the Hospital Anxiety and Depression Symptom scale (HADS) [24].

Sex was encoded as a dummy variable with value 0 for male and 1 for female. Age took an integer value ranging from 18 to 87. Those anxiety and depression scores with values above 10, the cut-off for asymptomatic, were excluded. This reduced the number of subject scores from 619 to 578 for anxiety and to 606 for depression.

All subsets regression [38] was used to select the "best" subset of the 20 PCs and the best rotation of the retained PCs. Let the 20 PCs be $\psi_i$, i=1,20 where $\psi_{i,k}$ is the value of $\psi_i$ for subject k, i.e. the projection of the original 158 measures for subject k onto $\psi_i$. BMDP9R was used to solve for the A's by minimizing the sum of the squared errors, the $\varepsilon_k$'s, in equation (4). The solution maximizes R, the multiple correlation of the dependent variable, e.g. age, with the retained $\psi_i$'s. For PCs which were eliminated, the corresponding $A_i$ was set to zero.

$$age_k = A_0 + \varepsilon_k + \sum_{i=1,20} \mathbf{A_i \psi_{i,k}} \quad (4)$$

# Results

The primary and secondary processing were completed for 619 members of the cohort for the resting recordings and for 617 for the task recordings. Although the thresholds for acceptance of significant differences are stringent, there are more than 380,000 results to report with $p < 10^{-8}$ for each, i.e. $p < 10^{-2}$ for each when corrected for multiple comparisons. The majority are for comparisons of the measures from individual subjects. The results are summarized in the tables which follow with several detailed examples presented.

# Identification of validated neuroelectric currents

The number of validated currents identified inside freesurfer ROI's for the resting recordings ($n$ = 619) was *mean* 6,365,040 per subject. This reduces to a *mean* of 455 simultaneously active currents at each moment throughout the recording. This number is lower than those for the subset

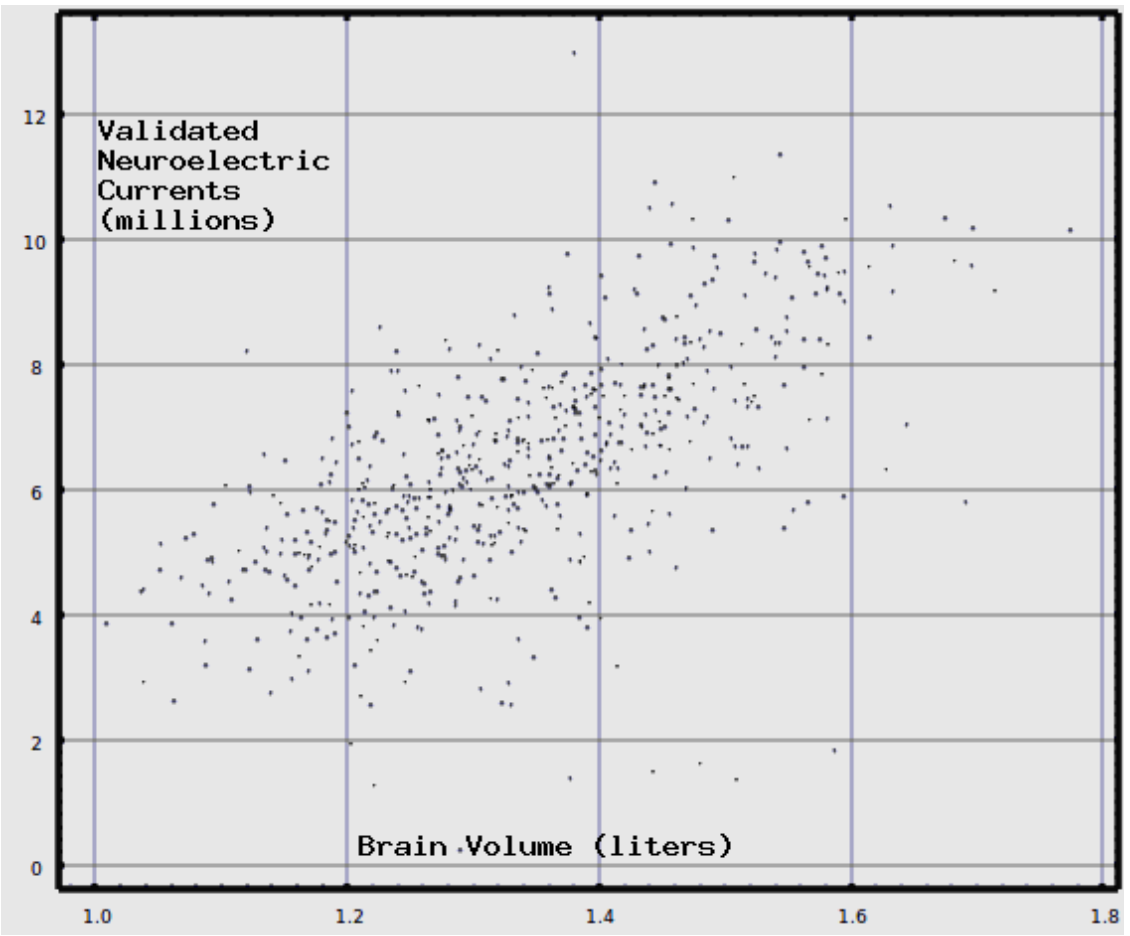

Figure 4. The brain volume for each subject was obtained using Freesurfer (see Figure 1). The count of validated neuroelectric sources shows the expected increase with brain volume. The Pearson's correlation between brain volume and current count is 0.634, $df = 617$, $p < 10^{-44}$

of the cohort aged 18-65, viz. 471 (rest) and 481 (task). The reduction is at least partially due to the progressive reduction in brain volume with age. Figure 4 shows the positive correlation between brain volume and the number of validated neuroelectric currents.

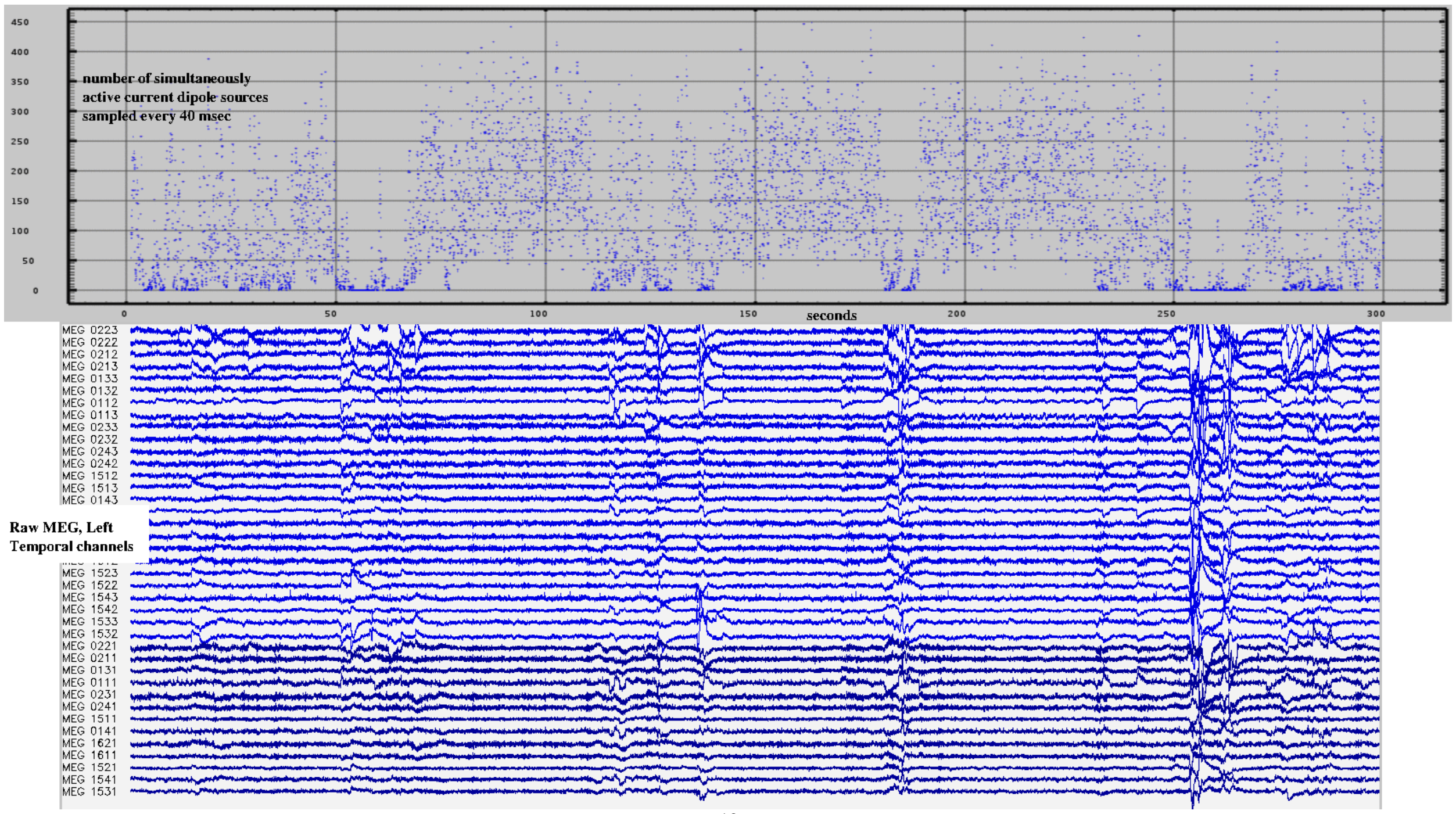

Figure 5: Each neuroelectric source is validated at $p < 10^{-12}$. The referee consensus solver automatically fails when the recordings are noisy. 300 seconds of raw MEG (lower) and neuroelectric current counts (upper) are shown. The number of validated ($p < 10^{-12}$) currents drops markedly when the MEG is noisy.

# Regional neuroelectric activity

Results of comparisons are reported for each subjects' ROIs and for the cohort as a whole. Sample results are shown for a single subject in Table 2 and in Figure 6. A significant difference in activity between rest and task was seen for a *mean* of 59.3% of the 158 ROIs for each subject, i.e. 57,803 of 97,486 ROIs. The threshold for significance for each test was $p < 10^{-8}$, i.e. $p < 10^{-2}$ for each when corrected for multiple comparisons.

29,007 of the tests demonstrated greater activity during task than rest; 28,796 demonstrated the opposite. This insignificant difference was reflected in the same comparisons for each ROI, i.e. there was no ROI for which the preponderance of subjects demonstrated either greater or lesser activity in one state vs the other, even with $10^{-2}$ used as the threshold for significance.

| Rest | | Task | | Rest vs Task | | |
|---|---|---|---|---|---|---|
| Count | Activity | Count | Activity | $\chi^2$ (*df*=1) | | ROI |
| 46675 | 2.41695 | 42270 | 2.22041 | -159.716 | $p < 10^{-35}$ | R_Hippocampus |
| 16733 | 2.28240 | 15312 | 2.11868 | -44.333 | $p < 10^{-10}$ | R_Pallidum |
| 55022 | 2.50938 | 50363 | 2.33002 | -144.721 | $p < 10^{-32}$ | R_Putamen |
| 44225 | 1.27562 | 41630 | 1.21808 | -45.703 | $p < 10^{-10}$ | R_Thalamus-Proper |
| 207291 | 1.68703 | 185400 | 1.53062 | -927.580 | $p < 10^{-203}$ | R_UnsegmentedWhiteMatter |
| 17808 | 1.75472 | 16847 | 1.68396 | -14.672 | $p < 10^{-3}$ | L_ctx_bankssts |
| 6298 | 1.31083 | 4815 | 1.01662 | -177.416 | $p < 10^{-39}$ | L_ctx_caudalanteriorcingulate |
| 40347 | 1.37955 | 45897 | 1.49210 | 136.668 | $p < 10^{-30}$ | L_ctx_caudalmiddlefrontal |
| 16162 | 1.48261 | 14751 | 1.37269 | -45.801 | $p < 10^{-10}$ | L_wm_bankssts |
| 13732 | 1.08158 | 15246 | 1.21814 | 102.223 | $p < 10^{-23}$ | L_wm_caudalanteriorcingulate |
| 44136 | 1.47019 | 44964 | 1.51937 | 24.115 | $p < 10^{-6}$ | L_wm_caudalmiddlefrontal |

Table 2. Typical regional counts and activity measures are shown for a sampling of ROIs for this individual. The white matter ROIs in the 3rd block are adjacent to the cortical ROIs in the 2nd block. Note that the normalized activity measure is 1.0 if the number of currents validated within an ROI is precisely that ROI's fair share based on its volume. Hence the sub-cortical ROIs demonstrate relatively high activity for this individual. The significance of the change in activity from rest to task is shown in the 6th column. The $\chi^2$ statistic (column 5) is positive/negative when task activity is greater/less than rest activity.

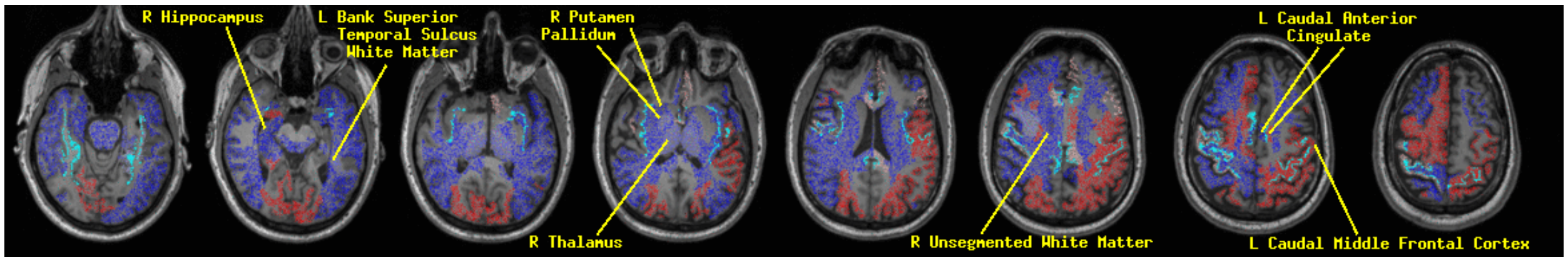

Figure 6. Differences are shown for the same subject presented in Table 2. Individual ROIs whose activity during task performance was greater (red) or less (blue) than activity during rest are indicated, $p < 10^{-8}$ for each. Color is fully saturated for $p < 10^{-12}$. This is the same subject whose counts and measures are shown in Table 2. The labelled ROIs are those listed in the table. Borders between cortex and white matter for 4 ROIs are shown as landmarks: precentral, cingulate, insula and fusiform.

# Network connectivity

The number of current pairs which occur simultaneously much more often than expected was counted, $p < 10^{-8}$ for each pair, $p < 0.05$ for each when corrected for multiple comparisons. For each individual, the number of instances of such pairs had a *mean* = 1,000,008, i.e. this count is about 1/6th the count for the validated neuroelectric currents. Sample results are shown for a single subject in Table 3.

Figure 7 shows the number of pair instances vs the pair separation for a single subject. For pairs of currents with relatively short separation, e.g. a few cm, it is possible that some form of cross-correlation analysis of the current waveforms may enable identification of pairs which fall on the same neuronal transmission line. Marrying that identification with each pair's localization and white matter tractography may produce realistic estimates of action potential propagation velocities.

| Count | Connectivity | ROI |
|---|---|---|
| 16,377 | 4.9448 | R_Hippocampus |
| 2,381 | 1.88862 | R_Pallidum |
| 9,253 | 2.46307 | R_Putamen |
| 8,927 | 1.50517 | R_Thalamus-Proper |
| 29,331 | 1.42010 | R_UnsegmentedWhiteMatter |
| 2,074 | 1.19306 | L_ctx_bankssts |
| 355 | 0.43091 | L_ctx_caudalanteriorcingulate |
| 9,236 | 1.74554 | L_ctx_caudalmiddlefrontal |
| 958 | 0.51305 | L_wm_bankssts |
| 384 | 0.17638 | L_wm_caudalanteriorcingulate |
| 5,630 | 1.10577 | L_wm_caudalmiddlefrontal |

Table 3. Typical regional pair counts and connectivity measures from the resting recordings are shown for the same ROIs and individual as those in Table 2. Note that as for activity, the normalized connectivity measure is 1.0 if the number of validated pairs for which one leg falls within a ROI is precisely that ROI's fair share of the total based on its volume. As for activity, the sub-cortical ROIs demonstrate relatively high connectivity for this individual.

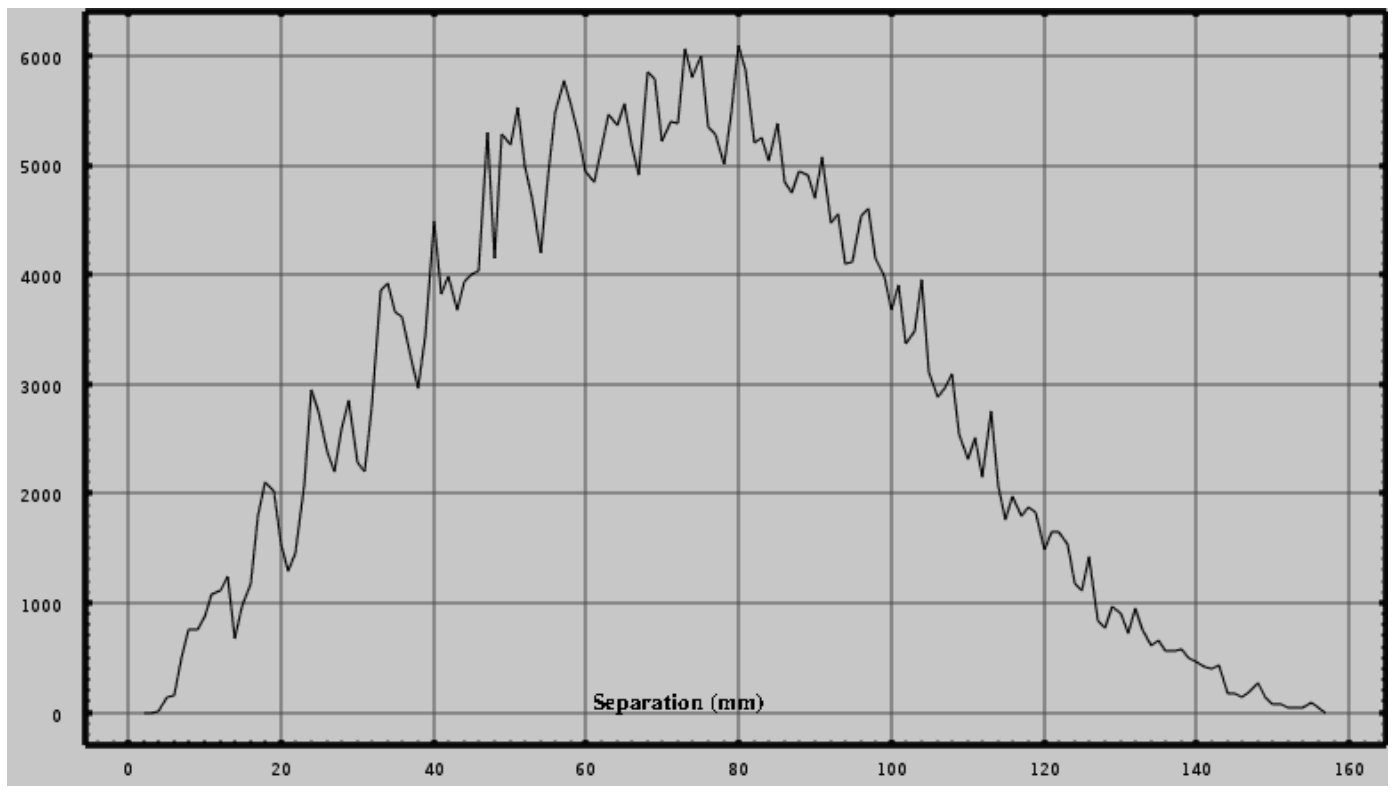

Figure 7. The number of pair instances validated at $p < 10^{-8}$ (y-axis) is shown as a function of pair separation (x-axis). These data are from the resting recordings from the same subject as shown in Tables 2 and 3 and Figure 6.

# Cortex vs adjacent white matter – cortical excitability

Comparisons are reported for each cortical/white matter ROI pair for individuals and for the cohort as a whole. Sample results are shown for a single subject in Tables 4 and 5 and Figure 8.

| Activity | | Cortex vs Adjacent White Matter | | |
|---|---|---|---|---|
| Cortex | White matter | $\chi^2$ ($df$=1) | | ROI |
| 1.75472 | 1.48261 | 240.723 | $p < 10^{-53}$ | L_bankssts |
| 1.31083 | 1.08158 | 192.226 | $p < 10^{-42}$ | L_caudalanteriorcingulate |
| 1.37955 | 1.47019 | -88.253 | $p < 10^{-20}$ | L_caudalmiddlefrontal |

Table 4. Typical regional pair comparisons of cortical vs white matter activity during rest are shown for the same individual as shown in Tables 2 and 3. The ratio of the volumes of the two ROIs in a pair is used to weight the expected count ratio. When $\chi^2$ is greater/less than 0, the cortical activity is greater/less than the white matter activity.

| Activity: Rest | | Activity: Task | | Rest → Task: cortex vs adjacent white matter | | |
|---|---|---|---|---|---|---|
| Cortex | White matter | Cortex | White matter | $\chi^2$ ($df$=3) | | ROI |
| 1.75472 | 1.48261 | 1.68396 | 1.37269 | 5.250 | -- | L_bankssts |
| 1.31083 | 1.08158 | 1.01662 | 1.21814 | -276.980 | $p < 10^{-58}$ | L_caudalanteriorcingulate |
| 1.37955 | 1.47019 | 1.49210 | 1.51937 | 23.046 | -- | L_caudalmiddlefrontal |

Table 5. Typical change from rest to task for cortex vs white matter regional pairs are shown for the same individual as shown in Tables 2-4. The ratio of the volumes of the two ROIs in a pair and the ratio of the total counts during rest and task are used to weight the expected count ratios. When $\chi^2$ is greater/less than zero, cortical vs white matter activity increased/decreased during task compared with rest.

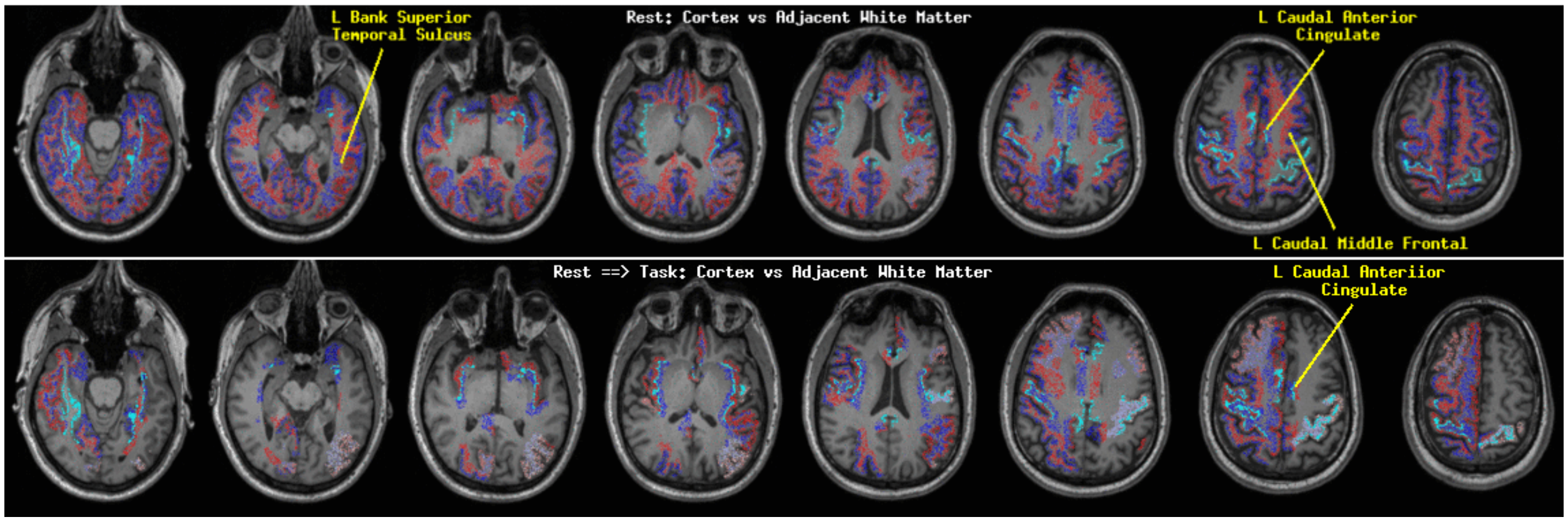

Figure 8. Differences are shown for the same subject presented in Tables 2-5 and figure 6. **Upper Panel:** Differences between the activity in cortical and adjacent white matter ROI pairs are shown. Each pair which shows a difference, $p < 10^{-8}$, is indicated in color with the ROI with higher activity shown in red, the adjacent ROI in blue. Color is fully saturated for $p < 10^{-12}$. **Lower Panel:** The change from rest to task in the difference between cortex and adjacent white matter is shown, $p < 10^{-8}$. For each pair, cortex vs white matter activity changed in the indicated direction for task compared with rest. E.g. for the left caudal anterior cingulate, the cortex was significantly more active (red) than the adjacent white matter (blue) during rest (upper panel). The activity in the cortex dropped significantly (blue) compared with the adjacent white matter (red) during the task (lower panel).

For the resting recordings for the subjects aged 18-65 ($n$ = 411), 23,008 of the cortex/white matter ROI pairs demonstrated a significant difference in current density at $p < 10^{-8}$. That is 82.3% of all the ROI pairs. For those which showed significance, the white matter ROI showed higher current density than the adjacent cortical ROI for 62.7% (14,438 pairs), i.e. the activity in the white matter was higher than that in the cortex in the preponderance of the ROIs, $\chi^2 = 1496$, $df = 1$, $p < 10^{-323}$.

The 32 single ROIs for which the preponderance of subjects demonstrated a cortex vs adjacent white matter difference in one direction ($p < 10^{-8}$ for each) are listed in Table 6 and are graphically presented in Figure 9. The table shows that the cortex is consistently more active than the adjacent white matter for 2 ROI pairs, left and right inferior parietal. The white matter is more consistently active than the adjacent cortex for 30 ROI pairs.

| Cortex > WM | Cortex < WM | $\chi^2$ (df=1) | | ROI |
|---|---|---|---|---|
| 425 | 109 | 186.996 | $p < 10^{-41}$ | L_inferiorparietal |
| 385 | 154 | 99.000 | $p < 10^{-22}$ | R_inferiorparietal |
| 332 | 185 | 41.796 | $p < 10^{-9}$ | L_parsopercularis |
| 330 | 185 | 40.824 | $p < 10^{-9}$ | R_parsopercularis |
| 145 | 341 | -79.044 | $p < 10^{-18}$ | L_cuneus |
| 54 | 177 | -65.492 | $p < 10^{-15}$ | L_frontalpole |
| 169 | 356 | -66.606 | $p < 10^{-15}$ | L_inferiortemporal |
| 186 | 364 | -57.606 | $p < 10^{-13}$ | L_insula |
| 187 | 359 | -54.182 | $p < 10^{-12}$ | L_isthmuscingulate |
| 95 | 411 | -197.342 | $p < 10^{-44}$ | L_lingual |
| 189 | 319 | -33.266 | $p < 10^{-8}$ | L_medialorbitofrontal |
| 65 | 487 | -322.614 | $p < 10^{-71}$ | L_middletemporal |
| 116 | 413 | -166.746 | $p < 10^{-37}$ | L_paracentral |
| 152 | 303 | -50.112 | $p < 10^{-11}$ | L_parahippocampal |
| 127 | 362 | -112.934 | $p < 10^{-25}$ | L_pericalcarine |
| 182 | 321 | -38.410 | $p < 10^{-9}$ | L_postcentral |
| 181 | 329 | -42.948 | $p < 10^{-10}$ | L_posteriorcingulate |
| 70 | 498 | -322.506 | $p < 10^{-71}$ | L_precuneus |
| 142 | 411 | -130.850 | $p < 10^{-29}$ | L_rostralmiddlefrontal |
| 131 | 438 | -165.638 | $p < 10^{-37}$ | L_superiorfrontal |
| 173 | 355 | -62.734 | $p < 10^{-14}$ | L_superiorparietal |
| 146 | 348 | -82.598 | $p < 10^{-18}$ | L_superiortemporal |
| 155 | 324 | -59.626 | $p < 10^{-13}$ | R_cuneus |
| 106 | 401 | -171.646 | $p < 10^{-38}$ | R_entorhinal |
| 65 | 176 | -51.124 | $p < 10^{-12}$ | R_frontalpole |
| 155 | 394 | -104.044 | $p < 10^{-23}$ | R_insula |
| 168 | 363 | -71.610 | $p < 10^{-16}$ | R_isthmuscingulate |
| 171 | 343 | -57.556 | $p < 10^{-13}$ | R_lateraloccipital |
| 131 | 349 | -99.008 | $p < 10^{-22}$ | R_lingual |
| 64 | 486 | -323.788 | $p < 10^{-71}$ | R_middletemporal |
| 118 | 432 | -179.264 | $p < 10^{-40}$ | R_paracentral |
| 171 | 302 | -36.280 | $p < 10^{-8}$ | R_parahippocampal |
| 113 | 351 | -122.076 | $p < 10^{-27}$ | R_pericalcarine |
| 176 | 324 | -43.808 | $p < 10^{-10}$ | R_postcentral |
| 181 | 330 | -43.446 | $p < 10^{-10}$ | R_posteriorcingulate |
| 40 | 529 | -420.246 | $p < 10^{-92}$ | R_precuneus |
| 125 | 433 | -170.006 | $p < 10^{-38}$ | R_rostralmiddlefrontal |
| 126 | 437 | -171.794 | $p < 10^{-38}$ | R_superiorfrontal |
| 167 | 349 | -64.192 | $p < 10^{-14}$ | R_superiorparietal |
| 138 | 371 | -106.658 | $p < 10^{-24}$ | R_superiortemporal |

Table 6. Cohort-wide differences are shown. These are ROIs during rest for which the preponderance of subjects ($p < 10^{-8}$) demonstrated a cortical vs adjacent white matter activity difference in one direction. For the ROIs whose $\chi^2$ statistic is negative (unshaded), the cortex was less active than the adjacent white matter for the preponderance of subjects for whom there was a significant difference, $p < 10^{-8}$ for each subject. The ROIs listed in the table are the same as those shown in color in Figure 9. The numbers of subjects for whom the cortex differed from the white matter in the indicated direction are shown in columns 1 and 2, $p < 10^{-8}$ for each of the enumerated subjects.

For the 410 subjects aged 18-65 with both resting and task recordings, resting cortex vs white matter activity was different from task cortex vs white matter activity for 12,132 ROI pairs. That is 43.5% of all the ROIs. There was no preponderance of ROI pairs which changed in one direction or the other. And there was no single ROI pair for which the preponderance of subjects demonstrated a significant change in either direction, even at $p < 10^{-3}$. That is consistent with the fact that there were 68 comparisons so the chance that one would reach the threshold of $10^{-3}$ is only 1/14.

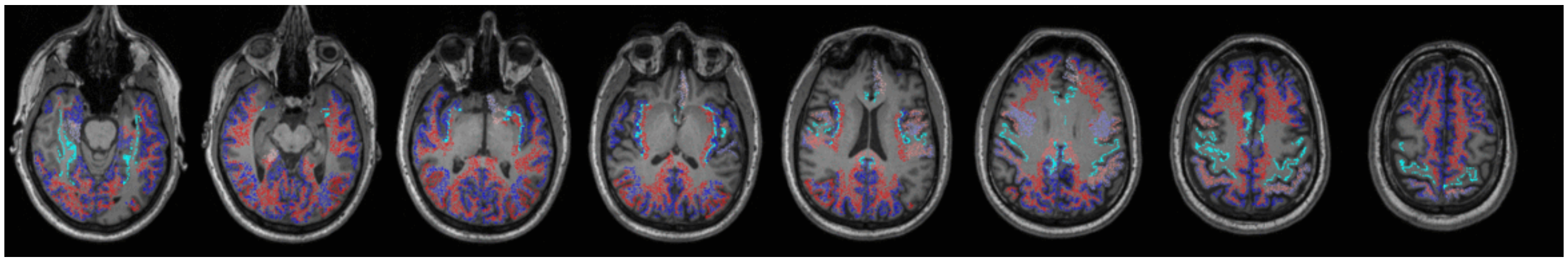

Figure 9. Cohort-wide differences are shown. For each pair of ROIs shown in color, the preponderance of resting recordings ($p < 10^{-8}$) demonstrated cortex vs adjacent white matter activity difference in the indicated direction, i.e. the red ROI was more active than the adjacent blue ROI, $p < 10^{-8}$ for each. For fully saturated colors, $p < 10^{-12}$. The ROIs are listed in Table 6.

# Activity vs connectivity -- locality

Comparisons of connectivity with activity are reported for each ROI for individuals and for the cohort as a whole. The cohort wide group results are confined to those aged 18-65. Sample results are shown for a single subject in Tables 7 and 8 and Figure 10.

| | | | | Resting Connectivity vs Activity | | |
|---|---|---|---|---|---|---|
| Count | Connectivity | Count | Activity | $\chi^2$ (*df*=1) | | ROI |
| 16377 | 4.94448 | 46675 | 2.416965 | 999.900 | $p < 10^{-218}$ | R_Hippocampus |
| 2381 | 1.88862 | 16733 | 2.28240 | -159.709 | $p < 10^{-35}$ | R_Pallidum |
| 9253 | 2.46307 | 55022 | 2.50938 | -5.539 | --- | R_Putamen |
| 8927 | 1.50517 | 44225 | 1.27562 | 384.668 | $p < 10^{-84}$ | R_Thalamus-Proper |
| 29331 | 1.42010 | 207291 | 1.68703 | -999.900 | $p < 10^{-218}$ | R_UnsegmentedWhiteMatter |
| 2074 | 1.19306 | 17808 | 1.75472 | -636.055 | $p < 10^{-139}$ | L_ctx_bankssts |
| 355 | 0.43091 | 6298 | 1.31083 | -999.900 | $p < 10^{-218}$ | L_ctx_caudalanteriorcingulate |
| 9236 | 1.74554 | 43047 | 1.37955 | 782.128 | $p < 10^{-171}$ | L_ctx_caudalmiddlefrontal |
| 958 | 0.51305 | 16162 | 1.48261 | -999.900 | $p < 10^{-218}$ | L_wm_bankssts |
| 384 | 0.17638 | 13732 | 1.08158 | -999.900 | $p < 10^{-218}$ | L_wm_caudalanteriorcingulate |
| 5630 | 1.10577 | 44136 | 1.47019 | -905.090 | $p < 10^{-198}$ | L_wm_caudalmiddlefrontal |

Table 7. Typical regional connectivity and activity measures are shown for a sampling of ROIs for this individual during rest. The white matter ROIs in the 3rd block are adjacent to the cortical ROIs in the 2nd block. Note that the normalized measures are 1.0 if the number of pairs or currents validated within a ROI is precisely that ROI's fair share based on its volume. Hence the sub-cortical ROIs demonstrate relatively high activity for this individual. The significance of the difference in connectivity vs activity is shown in the 6th column. The $\chi^2$ statistic (column 5) is positive/negative when connectivity is greater/less than activity.

For the resting recordings, the activity measure differed from connectivity for 82,918 of 97,802 ROIs, $p < 10^{-8}$ for each, i.e. 84.8%. For those comparisons which showed differences, the activity measure was greater than the connectivity measure for 58.7% (48,652 ROIs), i.e. the activity measure was greater than the connectivity measure in the preponderance of the ROIs, $\chi^2 = 2495.92$, $df = 1$, $p < 10^{-323}$.

| Rest | | Task | | Rest → Task: connectivity vs activity | | |
|---|---|---|---|---|---|---|
| Connectivity | Activity | Connectivity | Activity | $\chi^2$ (*df*=1) | | ROI |
| 4.94448 | 2.41695 | 1.64761 | 2.22041 | -2881.228 | $p < 10^{-323}$ | R_Hippocampus |
| 1.88862 | 2.28240 | 2.22257 | 2.11868 | 181.279 | $p < 10^{-40}$ | R_Pallidum |
| 2.46307 | 2.50938 | 1.82803 | 2.33002 | -17.520 | --- | R_Putamen |
| 1.50517 | 1.27562 | 1.34027 | 1.21808 | 26.994 | --- | R_Thalamus-Proper |
| 1.42010 | 1.68703 | 1.41384 | 1.53062 | 831.712 | $p < 10^{-182}$ | R_UnsegmentedWhiteMatter |
| 1.19306 | 1.75472 | 3.02629 | 1.68396 | 1771.912 | $p < 10^{-323}$ | L_ctx_bankssts |
| 0.43091 | 1.31083 | 0.06222 | 1.10662 | -138.270 | $p < 10^{-31}$ | L_ctx_caudalanteriorcingulate |
| 1.74554 | 1.37955 | 1.38375 | 1.49210 | -89.333 | $p < 10^{-20}$ | L_ctx_caudalmiddlefrontal |
| 0.51305 | 1.48261 | 0.15680 | 1.37269 | -234.933 | $p < 10^{-52}$ | L_wm_bankssts |
| 0.17638 | 1.08158 | 0.36038 | 1.21814 | 152.520 | $p < 10^{-34}$ | L_wm_caudalanteriorcingulate |
| 1.10577 | 1.47019 | 1.89572 | 1.51937 | 1430.320 | $p < 10^{-312}$ | L_wm_caudalmiddlefrontal |

Table 8. Typical changes from rest to task for connectivity vs activity measures are shown for the same individual as shown in Tables 2-6,7 above. When $\chi^2$ is greater/less than 0, the connectivity measure increased/decreased compared with the activity measure during task compared with rest.

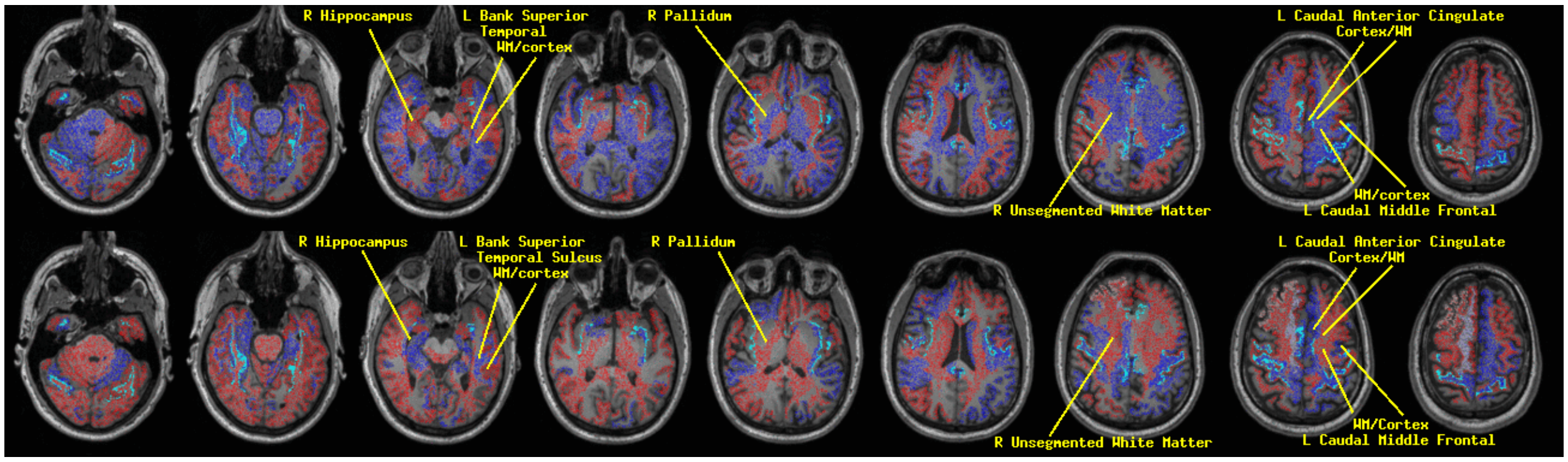

Figure 10. Differences are shown for the same subject presented in Tables 2-7 and Figure 6-9. **Upper Panel:** Differences between connectivity and activity are shown. Each ROI which shows a difference, $p < 10^{-8}$, is indicated in color. ROIs with higher connectivity than activity are shown in red; ROIs with lower connectivity than activity are shown in blue. Color is fully saturated for $p < 10^{-12}$. **Lower Panel:** The change from rest to task in the difference between connectivity and activity is shown, $p < 10^{-8}$. For each ROI, connectivity vs activity changed in the indicated direction for task compared with rest. E.g. for the left caudal anterior cingulate cortex, connectivity vs activity dropped during task compared with rest (blue). It's interesting that the adjacent white matter (WM) changed in the opposite direction (red).

The 73 ROIs for which the preponderance of subjects demonstrated connectivity vs activity difference in one direction ($p < 10^{-8}$ for each) are listed in Table 9 and are graphically presented in Figure 11. The table shows that connectivity is consistently greater than activity for 12 ROIs and consistently less than activity for 61.

For all 158 ROIs, resting connectivity vs activity was different from task connectivity vs activity for 77,707 ROIs. That is 79.4% of all the ROIs. For those which showed a significant change, the connectivity measure increased compared with the activity measure in 51.8%, a small but significant preponderance, $\chi^2 = 71$, $df = 1$, $p < 10^{-14}$. There was no single ROI for which the preponderance of subjects demonstrated a significant change in either direction, even at $p < 10^{-3}$. That is consistent with the fact that there were 158 comparisons so the chance that even one would reach the threshold of $10^{-3}$ is only 1/6.

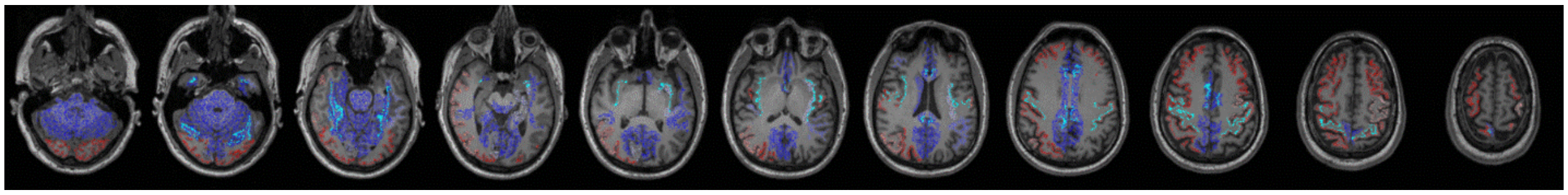

Figure 11. Cohort-wide differences are shown. For each ROI shown in color, the preponderance ($p < 10^{-8}$) of subjects demonstrated a connectivity vs activity difference in the indicated direction. Red indicates that connectivity was greater than activity for the ROI for the preponderance of the subjects for whom there was a significant difference between connectivity and activity, $p < 10^{-8}$ for each subject. For fully saturated colors, $p < 10^{-12}$. The ROIs are listed in Table 9.

| Connectivity > Activity | Connectivity < Activity | $\chi^2$ (df=1) | | ROI |
|---|---|---|---|---|
| 354 | 210 | 36.764 | $p < 10^{-8}$ | L_ctx_caudalmiddlefrontal |
| 411 | 167 | 103.002 | $p < 10^{-23}$ | L_ctx_lateraloccipital |
| 365 | 204 | 45.554 | $p < 10^{-10}$ | L_ctx_precentral |
| 393 | 184 | 75.702 | $p < 10^{-17}$ | L_ctx_rostralmiddlefrontal |
| 368 | 190 | 56.780 | $p < 10^{-13}$ | R_ctx_caudalmiddlefrontal |
| 354 | 203 | 40.934 | $p < 10^{-9}$ | R_ctx_inferiorparietal |
| 419 | 166 | 109.416 | $p < 10^{-24}$ | R_ctx_lateraloccipital |
| 362 | 209 | 40.996 | $p < 10^{-9}$ | R_ctx_middletemporal |
| 367 | 198 | 50.559 | $p < 10^{-11}$ | R_ctx_postcentral |
| 378 | 197 | 56.974 | $p < 10^{-13}$ | R_ctx_precentral |
| 389 | 184 | 73.342 | $p < 10^{-17}$ | R_ctx_rostralmiddlefrontal |
| 381 | 191 | 63.110 | $p < 10^{-14}$ | R_ctx_superiorparietal |
| 82 | 512 | -311.278 | $p < 10^{-68}$ | Brain-Stem |
| 80 | 524 | -326.384 | $p < 10^{-72}$ | L_Cerebellum-Cortex |
| 94 | 493 | -271.210 | $p < 10^{-60}$ | L_Cerebellum-White-Matter |
| 216 | 363 | -37.320 | $p < 10^{-8}$ | L_Hippocampus |
| 182 | 343 | -49.372 | $p < 10^{-11}$ | R_Caudate |
| 88 | 514 | -301.454 | $p < 10^{-66}$ | R_Cerebellum-Cortex |
| 83 | 508 | -305.626 | $p < 10^{-67}$ | R_Cerebellum-White-Matter |
| 110 | 245 | -51.338 | $p < 10^{-12}$ | L_ctx_caudalateriorcingulate |
| 151 | 327 | -64.802 | $p < 10^{-15}$ | L_ctx_cuneus |
| 145 | 435 | -145.000 | $p < 10^{-32}$ | L_ctx_fusiform |
| 132 | 273 | -49.088 | $p < 10^{-11}$ | L_ctx_isthmuscingulate |
| 126 | 446 | -179.020 | $p < 10^{-40}$ | L_ctx_lingual |
| 190 | 356 | -50.468 | $p < 10^{-11}$ | L_ctx_medialorbitofrontal |
| 135 | 294 | -58.930 | $p < 10^{-13}$ | L_ctx_paracentral |
| 199 | 357 | -44.898 | $p < 10^{-10}$ | L_ctx_parahippocampal |
| 150 | 330 | 67.500 | $p < 10^{-15}$ | L_ctx_pericalcarine |
| 124 | 296 | -70.438 | $p < 10^{-16}$ | L_ctx_posteriorcingulate |
| 158 | 376 | -88.996 | $p < 10^{-20}$ | L_ctx_precuneus |
| 162 | 322 | -52.892 | $p < 10^{-12}$ | L_ctx_rostralanteriorcingulate |
| 161 | 296 | -39.878 | $p < 10^{-9}$ | L_ctx_transversetemporal |
| 198 | 336 | -35.662 | $p < 10^{-8}$ | R_ctx_bankssts |
| 114 | 251 | -51.420 | $p < 10^{-12}$ | R_ctx_caudalanteriorcingulate |
| 182 | 311 | -33.754 | $p < 10^{-8}$ | R_ctx_cuneus |

Table 9. These are ROIs during rest for which the preponderance ($p < 10^{-8}$) of subjects demonstrated a connectivity vs activity difference in one direction. For the ROIs whose $\chi^2$ statistic is negative (unshaded), connectivity was less than activity for the preponderance of subjects for whom there was a significant difference between connectivity and activity, $p < 10^{-8}$ for each subject. The ROIs listed in the table are the same as those shown in color in Figure 11. The numbers of subjects for whom connectivity differed from activity in the indicated direction are shown in columns 1 and 2, $p < 10^{-8}$ for each of the enumerated subjects. Table 9 is continued below.

Table 9 is continued from above.

| Connectivity > Activity | Connectivity < Activity | $\chi^2$ (df=1) | | ROI |
|---|---|---|---|---|
| 192 | 360 | -51.130 | $p < 10^{-12}$ | R_ctx_entorhinal |
| 152 | 424 | -128.444 | $p < 10^{-29}$ | R_ctx_fusiform |
| 120 | 292 | -71.804 | $p < 10^{-16}$ | R_ctx_isthmuscingulate |
| 160 | 402 | -104.206 | $p < 10^{-23}$ | R_ctx_lingual |
| 157 | 376 | -89.982 | $p < 10^{-20}$ | R_ctx_medialorbitofrontal |
| 142 | 282 | -46.226 | $p < 10^{-10}$ | R_ctx_paracentral |
| 183 | 383 | -70.670 | $p < 10^{-16}$ | R_ctx_parahippocampal |
| 187 | 332 | -40.510 | $p < 10^{-9}$ | R_ctx_pericalcarine |
| 138 | 275 | -45.444 | $p < 10^{-10}$ | R_ctx_posteriorcingulate |
| 173 | 364 | -67.934 | $p < 10^{-15}$ | R_ctx_precuneus |
| 130 | 315 | -76.910 | $p < 10^{-17}$ | R_ctx_rostralanteriorcingulate |
| 166 | 303 | -40.018 | $p < 10^{-9}$ | R_ctx_transversetemporal |
| 183 | 375 | -66.064 | $p < 10^{-15}$ | L_wm_bankssts |
| 134 | 332 | -84.128 | $p < 10^{-19}$ | L_wm_caudalanteriorcingulate |
| 157 | 337 | -65.586 | $p < 10^{-15}$ | L_wm_cuneus |
| 159 | 403 | -105.934 | $p < 10^{-24}$ | L_wm_fusiform |
| 207 | 359 | -40.818 | $p < 10^{-9}$ | L_wm_inferiortemporal |
| 206 | 343 | -34.186 | $p < 10^{-8}$ | L_wm_insula |
| 131 | 440 | -167.215 | $p < 10^{-37}$ | L_wm_lingual |
| 162 | 331 | -57.932 | $p < 10^{-13}$ | L_wm_paracentral |
| 176 | 364 | -65.450 | $p < 10^{-15}$ | L_wm_parahippocampal |
| 197 | 355 | -45.224 | $p < 10^{-10}$ | L_wm_pericalcarine |
| 159 | 356 | -75.356 | $p < 10^{-17}$ | L_wm_posteriorcingulate |
| 171 | 388 | -84.236 | $p < 10^{-19}$ | L_wm_precuneus |
| 167 | 335 | -56.222 | $p < 10^{-13}$ | L_wm_rostralanteriorcingulate |
| 186 | 363 | -57.064 | $p < 10^{-13}$ | L_wm_superiortemporal |
| 202 | 357 | 42.978 | $p < 10^{-10}$ | L_wm_supramarginal |
| 143 | 275 | -41.684 | $p < 10^{-9}$ | L_wm_transversetemporal |
| 183 | 372 | -64.362 | $p < 10^{-14}$ | R_wm_bankssts |
| 140 | 313 | -66.068 | $p < 10^{-15}$ | R_wm_caudalanteriorcingulate |
| 181 | 394 | -78.902 | $p < 10^{-18}$ | R_wm_fusiform |
| 151 | 407 | -117.448 | $p < 10^{-26}$ | R_wm_lingual |
| 169 | 336 | -55.224 | $p < 10^{-12}$ | R_wm_paracentral |
| 172 | 376 | -75.940 | $p < 10^{-17}$ | R_wm_parahippocampal |
| 144 | 277 | -42.016 | $p < 10^{-10}$ | R_wm_parsorbitalis |
| 172 | 338 | -54.030 | $p < 10^{-12}$ | R_wm_posteriorcingulate |
| 171 | 383 | -81.126 | $p < 10^{-18}$ | R_wm_precuneus |
| 146 | 313 | -60.760 | $p < 10^{-14}$ | R_wm_rostralanteriorcingulate |

Table 9 is continued from above.

For the resting recordings for the full cohort, the Pearson's correlation between connectivity and activity is 0.63037. Although this is high, the percentage of ROIs for which there is a significant difference ($p < 10^{-8}$) between connectivity and activity is also high, 82,918 of the 97,802 ROIs, i.e. 84.8%. Furthermore, of those ROIs for which connectivity and activity differ, connectivity is less than activity for 58.7%, $\chi^2 = 2495.92$, $p < 10^{-323}$.

# Global activity patterns

A factor was identified with optimal correlation to each of several subject measures. The factor was a weighted sum of principle components obtained from the resting data regional activity measures for the entire cohort. The multiple correlation between factor scores on optimally selected global activity patterns and subject measures are listed in Table 10. A scatter plot of age vs factor score with regression line is shown in Figure 12. Figure 13 shows the ROIs which are the dominant contributors to the Age and Sex factors.

| | *R* | *F*-statistic | *n* | *p-value* |
|---|---|---|---|---|
| Age | 0.38883 | 8.99 | 12/606 | $p < 10^{-15}$ |
| Sex | 0.40362 | 7.82 | 15/603 | $p < 10^{-15}$ |
| Anxiety | 0.11756 | 4.03 | 2/575 | $p < 0.02$ |
| Depression | 0.16932 | 3.54 | 5/600 | $p < 0.004$ |
| ACE-R | 0.20621 | 3.87 | 7/610 | $p < 0.0004$ |
| PSQI | 0.14104 | 2.94 | 4/580 | $p < 0.02$ |

Table 10. Multiple correlation, *R,* between subject measures and factor scores on global activity patterns are listed. The relationships with Age and Sex are robust; those with anxiety and depression scores are much less so.

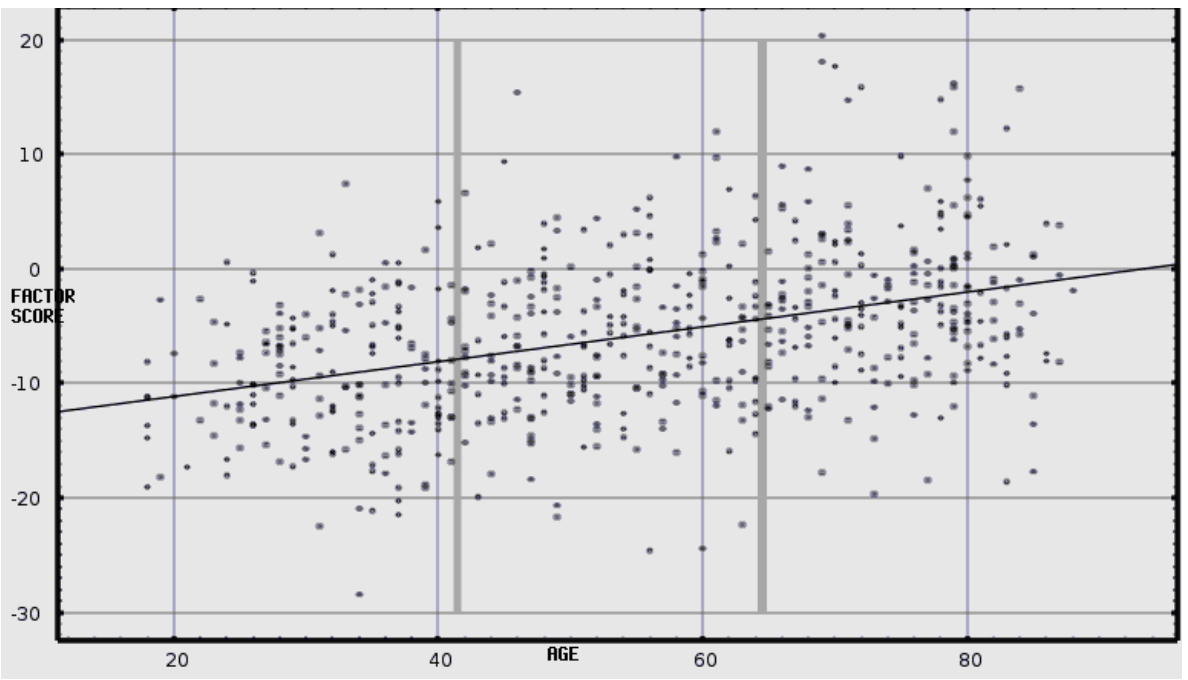

Figure 12. Age and score on the age factor (see text) are plotted on the x and y axes for each member of the CamCAN cohort. The best fit regression line is also shown; the multiple correlation from the factor rotation is 0.38883, Table 10.

The full cohort was divided by age into three subsets (vertical gray bars), ages 18-41, 42-64, and 65-87. The multiple correlations are 0.33985, 0.28373, and 0.33387 respectively, all with $p < 10^{-3}$. When the cohort is divided into male and female, $p < 10^{-5}$ for both subsets. Hence this result is stationary across the cohort.

The multiple correlation from the factor rotation for Sex vs factor score was 0.40362, Table 10. For the three data subsets, ages 18-41, 42-64, and 65-87, the multiple correlations are 0.47596, 0.45045, and 0.42335 respectively, all with $p < 10^{-5}$. The correlation between the Age factor and the Sex factor is 0.0012, i.e. these factors are independent. This was obtained by computing the dot product of the two factors.

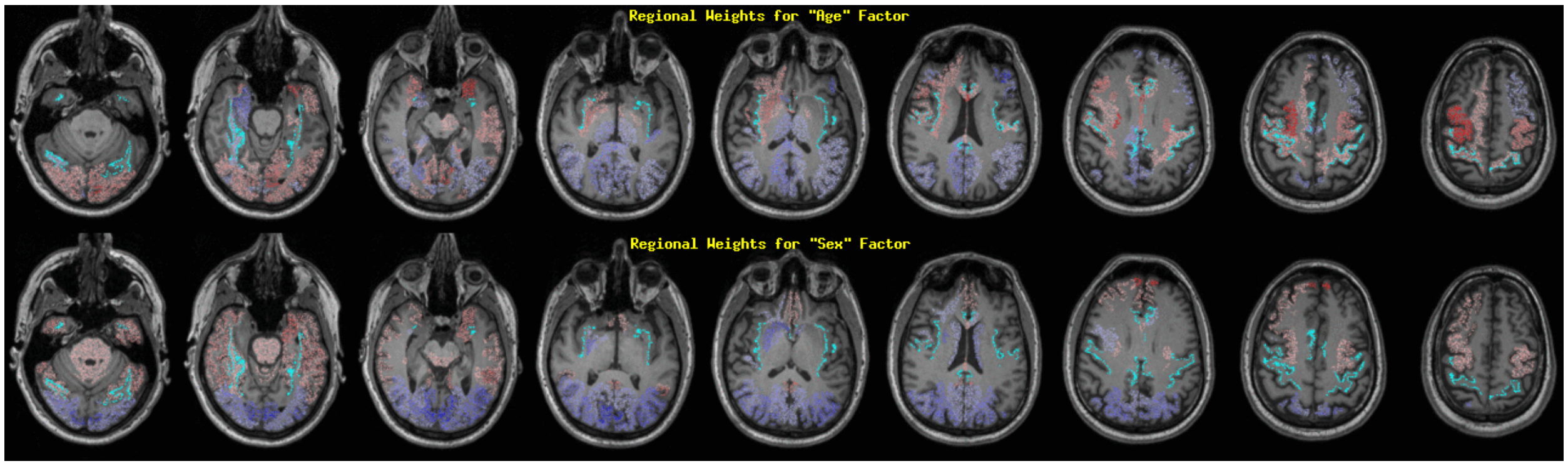

Figure 13. Regional weights are indicated in color. Red indicates a positive weight, blue a negative weight. More highly saturated colors indicate stronger weights. The weights were obtained using all subsets regression on the 20 "best" principle components of the regional activity measures (see text of Methods). **Upper Panel:** Weights are shown for the factor most highly correlated with age. **Lower Panel:** Weights are shown for the factor most highly correlated with sex. Note that the correlation between the two weight vectors is 0.0012, i.e. they are independent.

# Discussion

Figures 3 and 5 are illustrative examples which are included to provide an overall sense of the primary output of the referee consensus method, i.e. the copious but fragmentary validated neuroelectric currents. Figure 5 shows the number of simultaneously active currents for a 300-second interval from one recording. The number varies markedly with time throughout, as is typical. In particular, the number of validated currents drops precipitously during periods when the MEG recording contains movement and other artifacts. This property of the method is critically important, i.e. that the yield of the solver is sensitive to noise in the MEG recordings. It supports the supposition that (a) the probabilistic validation of the neuroelectric currents works and (b) that it automatically rejects noisy recordings, obviating the requirement to identify and remove noisy data segments from the analysis by hand.

For a two-minute period from one recording, Figure 3 illustrates all of the occurrences of validated currents for the 3000 locations which showed the greatest activity for this subject. The activity at each of the 3000 locations waxes and wanes markedly over the two minutes, as does the occurrence of simultaneous activity for pairs of locations. There are significant periods of activity at many locations during which there was little or no simultaneous pair activity.

There is considerable detectable neuroelectric activity from the white matter. This is not unprecedented. MEG-derived evoked responses from thalamocortical fibers have previously been reported [3,5]. The source magnetic fields are presumed to be due to synchronous volleys of action potentials, APs. Each AP produces a travelling current quadrupole. The approximate amplitude has been estimated at 100 Amp$^{-15}$ meters in an unmyelinated axon [39] with separation of 1 mm between the two dipoles forming the quadrupole assuming a propagation velocity of 1 meter/second. It is presumed that the velocity is greater in the myelinated fibers which comprise the white matter. Hence the velocity and dipole separation would be greater. This would decrease the distance-dependence of the magnetic field strength and so enhance the detectability of this activity. In addition the magnetic field due to an action potential in a single axon has been directly measured at about 150 x $10^{-12}$ Tesla [7].

A trivial explanation of the profuse findings reported here is that cortical activity is localized in nearby white matter due either to poor resolution or to head movements. The robust finding of differential activity between adjacent cortical and white matter ROIs argues against this. So too does the design of the method which relies on gradients between points within the brain that are 1 mm apart (Appendix 1) coupled with the use of once per second corrections to the forward solution using continuous head positioning information.

Under the assumption that the white matter is, in fact, the source of profuse measurable neuroelectric activity, the measured magnetic field components can only be due to synchronous volleys of APs (Table 11). These produce transient longitudinal intra-axonal currents which are nearly synchronous in many parallel running axons due to near simultaneous passage of propagating APs. The detected magnetic field waveforms, e.g. Figure 2, are envelopes which follow the high frequency waveforms of several AP volleys in sequence. The envelope of a single highly synchronized AP volley would require well under 10 msec to rise and fall. Hence this type of activity would be dominated by high frequency content. This is consistent with the observation that the yield of the solver improves when the low pass cut-off with which the signals are preprocessed is increased from 150 Hz to 330 Hz [33, figure 3]. It is also consistent with the observation from a typical task recording that the frequency content of the current waveforms includes profuse resonant activity with frequency content above 70 Hz [40].

| cortical ROI | | white matter ROI | effector |
|---|---|---|---|
| action potential volleys | | efferent action potential volleys | |
| ▲ | →→ | ▲ | Feed-forward excitation |
| ▼ | | ▼ | Feed-forward inhibition |
| population post-synaptic currents | | afferent action potential volleys | |
| ▲ | ←← | -- | Feed-forward excitation |
| ▼ | | -- | Feed-forward inhibition |

Table 11. The columns labelled "cortical ROI" and "white matter ROI" indicate the neuroelectric activity which is the presumed source of the MEG-derived activity measures. The effects of feed-forward excitation and inhibition on cortical and adjacent white matter activity are indicated with the up- and down- arrows (see text). → and ← indicate the flow direction for efferent (leaving) and afferent (arriving) action potentials (APs) at the cortical ROI. It is population post-synaptic currents (shaded) which are presumed to be the primary generators of the magnetic fields detected using standard MEG processing methods.

For the resting recordings for the full cohort, the Pearson's correlation between cortical and adjacent white matter activity is 0.81006. Although this is high as might be expected, the percentage of pairs for which there is a significant difference ($p < 10^{-8}$) between cortex and adjacent white matter is also high, 34,640 of the 42,092 ROI pairs, i.e. 82.3%. Furthermore, of those 34,640 ROI pairs which show a significant cortex vs white matter difference, the white matter shows more activity than the cortex for 62.6%, $\chi^2 = 2195$, $df = 1$, $p < 10^{-323}$.

The number of pairs for which the cortex vs adjacent white matter difference changes between resting and task recordings is also high, 19,402 of 42,092 pairs, 46.1%. But for these comparisons, there is no preponderance in the direction of the change. A simple speculative interpretation of this finding is that these rest vs task changes are due to task related feed-forward excitatory vs inhibitory afferent volleys arriving at the effected ROI pairs (Table 11). Feed-forward inhibition would reduce (a) the population post-synaptic currents (PPSCs), (b) cortical excitability, and therefore (c) the incidence of APs in cortical neurons. Efferent APs in the adjacent white matter would drop due to (c) but afferent APs would remain constant or rise; hence white matter vs cortical activity would go up. Note that it is cortical PPSCs which are presumed to be the primary generators of the magnetic fields detected using standard MEG processing methods.

Conversely, feed-forward excitation would produce the opposite of effects (a) – (c). Afferent activity in the white matter ROI would be comparable to that for feed-forward inhibition. Efferent volleys would increase since cortical APs would increase. But since both AP and PPSC activity in the cortex would increase, these would supersede increases in the white matter and cortical vs white matter activity would go up. Note that this implies that changes in cortical vs white matter activity provide a measure of changes in cortical excitability.

The initiation of an AP in a neuron produces a large voltage difference between the base of the dendritic tree and the AP trigger zone, i.e. across the neuron's cell body. Since the lumen of the cell body is much larger than the lumen of an axon, so too is the transient electric current across the soma and the resultant magnetic field. Hence the changes in AP activity within the cortex may have a predominant effect on changes in the difference between cortical and adjacent white matter activity.

The comparison of connectivity with activity provides a measure of locality of neuroelectric function, e.g. Tables 7, 8, 9, Figures 10, 11. It is not surprising that it is activity, i.e. locality, rather than connectivity which predominates for most ROIs.

There are robust findings for many ROIs for the preponderance of those in this large normative cohort from these connectivity vs activity comparisons (Figure 11, Table 9) and from the cortical vs adjacent white matter activity comparisons (Figure 9, Table 6). The strength of the statistical analysis provides considerable confidence in the veracity of the findings.

The principle components extracted from this large normative cohort may be conceived as normative patterns of regional neuroelectric activity. They may be used as measuring rods against which the MEG-derived activity measures of any individual may be compared. Factors composed of weighted sums of these patterns may be computed from the neuroelectric measures during rest from any cohort for which both MEG and symptoms scores or other clinically relevant measures are present. The factor scores may then be computed for the CamCAN cohort to identify the normal range. The regions and overall pattern for the factor may then elucidate the brain mechanisms which underlie the clinical entity.

# Concluding remarks

The primary processing of unaveraged MEG recordings with the referee consensus solver yielded approximately 10,000 neuroelectric current 80-msec waveforms per second localized with better than 5 mm spatial resolution, each validated at $p < 10^{-4}$ when corrected for multiple comparisons. The yield of the solver is summarized in Table 12.

For each cortical region, freesurfer identifies the adjacent white matter rim up to 5 mm thick. This dimension provides a nominal upper limit for the resolution demonstrated by the consistent finding of a significant difference in activity between cortical regions and the adjacent white

| | *n* | #currents/subject | #pairs/subject | #pair instances/subject |
|---|---|---|---|---|
| Rest | 619 | 6,365,040 | 66,190 | 1,000,008 |
| Task | 617 | 6,148,898 | 59,277 | 893,435 |

Table 12. The *mean* number of neuroelectric currents and number of validated neuroelectric pairs is shown. There were 516,264 comparison tests were performed for individual measures and 836 tests were performed for cohort-wide measures. $p < 10^{-12}$ was set as the threshold for acceptance of each test as significant. This insures that $p < 10^{-2}$ for each test for individual measures when corrected for multiple comparisons and $p < 10^{-5}$ for each test of cohort-wide measures when corrected for multiple comparisons. There were 386,189 significant tests for individual measures, i.e. a *mean* of 74.8% of all the comparisons for each subject. There were 221 significant tests of cohort-wide measures,

matter regions, Tables 4, 6 and Figures 8, 9.

The secondary processing used in this report reduced the voluminous set of fragmentary neuroelectric currents to regional counts over several minutes. Tests of differential activity using these tonic regional measures yield profuse results with robust significance.

(1) The referee consensus method provides profuse reliable measures of neuroelectric brain activity from cerebral cortex, white matter, and subcortical structures, e.g. hippocampus and cerebellum. In addition, robust high resolution measures of neuroelectric activity from the white matter are now accessible.
(2) Post-processing of validated neuroelectric currents using counting statistics provides profuse reliable measures of neuroelectric connectivity.
(3) Regional measures of activity and connectivity provide a statistically robust and detailed characterization of each individual's neuroelectric brain function. Comparisons of regional measures yield profuse and reliable differences between cortex and adjacent white matter ROIs, between connectivity and activity for a single ROI, between rest and task for a single ROI, etc.

(4) The group results are robust and demonstrate that a large minority of ROIs demonstrate stereotypic behavior whereas most ROIs vary from person to person. These findings may prove useful to (a) identify correspondences between these results and those found using standard MEG processing methods and/or fMRI and (b) to elucidate the mechanics of normal and abnormal brain function. Cohort-wide findings include (a) identification of specific ROIs whose behavior is similar for the preponderance of the cohort and (b) identification of patterns of brain activity which are correlated with age and sex. The same factor analytic approach used for these latter results may prove useful in identifying patterns of neuroelectric brain function related to disease states.

The differential comparison post-processing used here delivers profuse and detailed regional characterizations of the activity and connectivity for each individual. These findings are summarized in Table 13. They come with high confidence due to the statistical power of the individual results. This property provides great promise that these and other MEG-derived measures may prove useful for clinical diagnosis and for guidance of treatment.

| | *n* | differences | differences/subject (*mean*) | | differences cohort-wide | |
|---|---|---|---|---|---|---|
| Task → Rest: Activity | 617 | 57,803 | 93.7 of 158 | 59.3% | 0 of 158 | 0% |
| Rest: Activity vs Connectivity | 619 | 81,997 | 131.9 of 158 | 83.8% | 73 of 158 | 46.2% |
| Task: Activity vs Connectivity | 617 | 81,082 | 131.4 of 158 | 83.2% | 69 of 158 | 43.7% |
| Task → Rest: Activity vs Connectivity | 617 | 76,879 | 124.6 of 158 | 78.9% | 3 of 158 | 1.9% |
| Rest: Cortex vs Adjacent White Matter | 619 | 34,607 | 55.9 of 68 | 82.2% | 40 of 68 | 58.9% |
| Task: Cortex vs Adjacent White Matter | 617 | 34,513 | 55.9 of 68 | 82.2% | 36 of 68 | 52.9% |
| Task → Rest: Cortex vs White Matter | 617 | 19,308 | 32.3 of 68 | 46.0% | 0 of 68 | 0% |

Table 13. Summary of individual and cohort wide differential activity and connectivity comparisons. There were 516,264 comparison tests were performed for individual measures and 836 tests were performed for cohort-wide measures. $p < 10^{-8}$ was set as the threshold for acceptance of each test as significant. This insures that $p < 10^{-2}$ for each test for individual measures when corrected for multiple comparisons and $p < 10^{-5}$ for each test of cohort-wide measures when corrected for multiple comparisons. There were 386,189 significant tests for individual measures, i.e. a *mean* of 74.8% of all the comparisons for each subject. There were 221 significant tests of cohort-wide measures,

The profusion of significant findings for each individual and the implication that this provides a detailed characterization of regional neuroelectric brain function must be tested for repeat reliability. Follow up MEG studies have been obtained on 280 of the CamCAN cohort [22] and are being processed.

The neuroelectric waveforms are extracted with millisecond (msec) resolution. With each 40 msec step through a recording, the solver yields a *mean* of 455 simultaneously active currents, each with 80 msec duration and band pass of 10 – 250 Hz. Hence there is considerable opportunity to explore network connectivity with multivariate time series methods.

Although the content of the individual waveforms at frequencies below 10 Hz is strongly attenuated, the solver yields a new bolus of currents for each 40 msec step through the recording. Hence the count of currents throughout the brain is sampled at 25 Hz, providing access to low frequencies. Averaging of time series of these counts time locked to a trigger is one approach to accumulate sufficient counts to explore this low frequency range. Such averaged evoked responses could be accumulated over voxels up to 6 mm on a side [33, figures 9,10].

The results in this report depend on high localization resolution (better than 5 mm) of each neuroelectric current. That dependence is of particular importance in enabling comparison of activity measures between a cortical region and the adjacent white matter rim. Each current is a vector, i.e. it has a direction in space. Note that analysis of MEG provides two of the three direction vectors. The magnitude of the radial component of the current cannot be measured.

For the currents localized to cortical layers 4-6, that direction should be normal to the cortical surface and pointed in toward the white matter. For the currents localized to the deep white matter, that direction should be concordant with the prevailing direction of the axonal bundles' course. It may be possible to confirm this directional conjecture for the cortex using the MEG-derived direction of the currents and the measured geometry of the cortex from the anatomic MR imaging. Although not available for this cohort, it may be possible to confirm the directional conjecture for the white matter in a cohort for which both MEG and diffusion tensor imaging have been obtained.

A formidable problem posed by these results is that there is no comparable method or ground truth measures to which these findings may be compared. This is particularly true for the measures of activity from the white matter but only incrementally less so for the measures of connectivity. It may be that correspondences will be found to findings from fMRI and/or other MEG studies.

In order to promote exploration of these data by other workers, the output of the primary processing step is available from the authors. That output includes the complete population of neuroelectric currents from both resting and task recordings for the full cohort [http://stash.osgconnect.net/+krieger/ . Example files from this data set are contained in the supplementary file, S1_Tables.zip.] Access requires registration with CamCAN [http://www.mrc-cbu.cam.ac.uk/datasets/camcan/] which will also provide access if needed to the MR imaging, demographics, MEG recordings, and behavioral test results.

# Acknowledgements

We gratefully acknowledge the invaluable contributions made to this effort by the Cambridge (UK) Centre for Ageing and Neuroscience, the Extreme Science and Engineering Development Environment (XSede), the Open Science Grid (OSG), the San Diego Supercomputing Center, the Pittsburgh Supercomputing Center, the XSede Neuroscience Gateway, Darren Price, Mats Rynge, Rob Gardner, Frank Wurthwein, Derek Simmel, Mahidhar Tatineni, Ali Marie Shields. Data used in the preparation of this work were obtained from the CamCAN repository [22,23]. The OSG [41,42] is supported by the National Science Foundation, 1148698, and the US Department of Energy's Office of Science.

# Supporting Information

S1_Tables.zip: This archive contains several of the cohort-wide atlases and sample summary results from the individual presented in Tables 2-7 and Figures 6-10. It also contains a typical output file created by the referee consensus solver. This file contains the particulars from the 13,758 80-msec neuroelectric current waveforms identified in one second from this same individual's resting MEG recordings.

# Appendix I. The referee consensus method

The 2-fold task of the solver is (1) provide a robust measure of confidence that a dipole current is detected at location **X** and (2) estimate the time course of the current amplitude The solver is applied to one 80 msec data segment ($\mathbf{M}_{t=1,\ldots,80}$) at a time. A decision is made for one location at a time, e.g.: "Is there a dipole current present at location **X**?" To answer this question, spatial filters are constructed from the "viewpoints" for each of 90 distant "referee" locations distributed widely through the volume of the brain, e.g. **R**.

Filter $\mathbf{P}_{R!X'}$ is constructed with gain 1.0 at **R** and gain 0.0 at **X'** 1 mm from **X**. $\mathbf{P}_{R!X'}$ is applied to the 80 data vectors, $\mathbf{M}_{t=1,\ldots,80}$, to produce the 80-point univariate time series, $\mathbf{V}_{R!X'}$. A 2nd filter is constructed, $\mathbf{P}_{R!X}$, with gain 1.0 at **R** and gain 0.0 at **X**. $\mathbf{P}_{R!X}$ is also applied to $\mathbf{M}_{t=1,\ldots,80}$ to produce the 80-point univariate time series, $\mathbf{V}_{R!X}$. Note that there is a small contribution to $\mathbf{V}_{R!X'}$ from activity at **X** but none from **X'**. Contrariwise there is a small contribution to $\mathbf{V}_{R!X}$ from activity at **X'** but none from **X**. The difference filter is constructed, $\mathbf{P}_{R!X'-R!X}$. This has gain 0.0 at **R** and nearly equal and opposite gains at **X** and **X'**. $\mathbf{P}_{R!X'-R!X}$ applied to $\mathbf{M}_{t=1,\ldots,80}$ produces $\mathbf{V}_{R!X'-R!X}$, the difference: $\mathbf{V}_{R!X'} - \mathbf{V}_{R!X}$. Note that there is no contribution to this from **R**. Note too that each of these 3 filters is constructed with gain 0.0 at each of 89 other "referee" locations coarsely covering the brain so $\mathbf{V}_{R!X'-R!X}$ includes only small contributions from other neuroelectric currents. This insures that the primary contributors to $\mathbf{V}_{R!X'} - \mathbf{V}_{R!X}$ are currents close to **X** and/or **X'**.

The "opinion" from the viewpoint of referee **R** regards the presence of a current at **X** is obtained by evaluating this inequality:

$$(V_{R!X'-R!X} \bullet V_{R!X'})^2 > (V_{R!X'-R!X} \bullet V_{R!X})^2 \qquad (5)$$

If the inequality is true, then there is a current at **X** from the viewpoint of **R** since $\mathbf{V}_{R!X'}$ (left side) has no contribution from **X'**, $\mathbf{V}_{R!X}$ (right side) has none from **X**, and $\mathbf{V}_{R!X'-R!X}$ has nearly equal contributions from both.

This procedure is repeated for two vector components for each of the 90 referee locations to produce 180 yes/no "opinions." 114 or more must be "yes" ($p << 0.01$) to produce an acceptable "consensus" for this differential. The same procedure is repeated for each of the other 5 differentials since there are two differentials along each of the 3 spatial axes. Only if all 6 exceed the threshold, i.e. 57 or more of 90 for each of the 6, is a current accepted. $0.01^6 = 10^{-12}$ is therefore the threshold for accepting a current.

Once a location is validated, an eigenvector analysis is used to identify the 80-point time course of the current at that location as the waveform which captures the most variance in the complete set of $\mathbf{V}_{R!X'-R!X}$'s. The estimated signal/noise enhancement of 10 provided by this operation is detailed in the introduction. Note that the validation insures that there is a current present at **X** and not at any of the six **X'**s. Hence $\mathbf{V}_{R!X'-R!X}$ is used because the primary contributor to all of the $\mathbf{V}_{R!X'-R!X}$'s is due to the current at **X**. Because the current at **X** must dominate any current present at any of the **X'** s, the ability of the method to identify two currents near each other is limited to twice the distance, **X – X'**, i.e. 2 mm.

# Appendix II: The referee consensus method -- advantages

Like beamformers, the filters used in the referee consensus method are generated in sets, seven at a time. But unlike standard filter which are optimized to yield source space measures at the target/test location, **X**, all seven referee consensus filters are optimized to yield source space measures at *referee* locations remote from the test location, **X**.

Note that the estimated neuroelectric currents are two-dimensional vector quantities [9]. For the purpose of intelligibility, they are treated as scalars in the following explanation without loss of generalization.

For a particular referee location, **R**, all of the filters have gain 1.0 at **R. T**here is one filter constrained to have zero gain at **X**, filter $\mathbf{P}_{R!X}$. This filter, designated "**R not X**," is optimized to measure the signal at the referee location but with no contribution from activity at **X**. There are six other filters, each constrained to have zero gain at one of the six points 1.0 mm away from **X** along the x, y, or z-axis. Hence these filters are also optimized to measure the signal at the referee location but with no contribution from one of the locations 1 mm away from **X** along one of the coordinate axes.

These seven filters are used to generate six difference filters, e.g. $\mathbf{P}_{R!X'-R!X}$ where **X'** is the location +1 mm away from **X** along the x-axis.. Note that the difference filters are constrained to have zero gain at the referee location and near equal but opposite gains at **X** and the location 1 mm away. Hence they are optimized to measure the difference between the signals at **X** and at a location 1 mm away. The magnitude of the gain of these filters at these two "differential" locations is typically 0.05 – 0.08 [43, figure 6]. Note that the differencing delivers signal/noise enhancement of about 2.0 in measuring the differential between activity at **X** and activity 1 mm from **X**.

Conceptually this approach to source space measurement is upside down. The filters for the location at which measurements are made do not have gain 1.0 at the test location, **X**, as is the standard but rather have zero gain either at **X** or very near it. The power of the approach comes from the use of families of these filters to develop a consensus *decision* on the question: Is there or is there not a neurolectric current at the test location, **X**? For example, filters $\mathbf{P}_{R!X'}$, $\mathbf{P}_{R!X}$, and $\mathbf{P}_{R!X'-R!X}$ are each applied to an 80 msec MEG data segment to yield three 80 msec data traces, $\mathbf{V}_{R!X'}$, $\mathbf{V}_{R!X}$, and $\mathbf{V}_{R!X'-R!X}$. If $\mathbf{V}_{R!X'-R!X} \bullet \mathbf{V}_{R!X'} > \mathbf{V}_{R!X'-R!X} \bullet \mathbf{V}_{R!X}$, then there is a current at **X** from the "point of view" of referee **R**.

Note the number of MEG measures used to assess this inequality between two numbers, i.e. the two dot products. Each element of each of the **V**'s is the dot product of the corresponding filter with the 306 MEG measures for a single time point. Were both the filter weights and the MEG measures uncorrelated and normally distributed, the signal/noise enhancement due to the filtering operations would be approximately sqrt(306) ≈ 17 . To account for the certain significant departure from both assumptions, we divide the 306 degrees of freedom by 10 as a nominal correction to this estimate. Hence we estimate signal/noise enhancement due to the filters: sqrt(30) ≈ 5.

For the decision dot product, i.e. for evaluation of $\mathbf{V}_{R!X'-R!X} \bullet \mathbf{V}_{R!X'} > \mathbf{V}_{R!X'-R!X} \bullet \mathbf{V}_{R!X}$, each **V** is composed of 80 such filtered MEG measures. Hence the numbers $\mathbf{V}_{R!X'-R!X} \bullet \mathbf{V}_{R!X'}$ and $\mathbf{V}_{R!X'-R!X} \bullet \mathbf{V}_{R!X}$ are each obtained using 306 x 80 = 24,480 degrees of freedom. By the same reasoning as above with the same factor of 10 nominal correction, the estimated signal/noise enhancement due to the filters and the use of the 80-point times series is sqrt(2448) ≈ 50. Note that this estimate ignores the factor of 2.0 signal/noise enhancement inherent in the difference filter. Note too that this high signal/noise enhancement applies only to the numbers used in evaluating the inequality, not to the individual current amplitude measures, i.e. the elements of the **V**'s. This is why the emphasis in this paper is on the direct results of those decisions, i.e. the counts of validated neuroelectric currents.

There are 1080 binary decisions computed for a test location, **X**. one for each of two orthogonal orientations for each 90 *referee* locations for each of the six **X'** s. The collection of 2x90x6 = 1080 *decisions* provides a cost function which is used to assess the *consensus*: Is there or is there not a neuroelectric current at **X**. This procedure is robust enough to use $10^{-12}$ as the threshold p-value to accept **X** as a true source of a detectable magnetic field. If and only if this probabilistic threshold is met, the 1080 outputs of the difference filters, the $\mathbf{V}_{\mathbf{R!X'-R!X}}$**'s,** are combined by an eigenvector calculation which is akin to averaging [Appendix I] to generate an estimate of the 80-point time course for the current.

Computation for each of these decisions generates its own optimal difference filter, $\mathbf{P}_{\mathbf{R!X'-R!X}}$, and corresponding optimal 80-point current time series estimate, $\mathbf{V}_{\mathbf{R!X'-R!X}}$. As detailed above, the signal/noise enhancement estimate for the elements of the **V**'s, i.e. the current magnitude measures, are modest, ≈ 5, and likely comparable to those for most filtering methods. But the estimate obtained by this average-like operation over 1080 such difference filters again enhances the signal/noise by an additional factor of perhaps sqrt(1080/10) ≈ 10, producing total signal/noise enhancement for the current times series waveform estimate of about 50.